\documentclass[preprint,12pt]{elsarticle}

\usepackage{amssymb}
\usepackage{amsmath}
\usepackage{xcolor}
\usepackage[colorlinks=true,linkcolor=blue]{hyperref}
\usepackage{graphicx}
\usepackage{multicol}
\usepackage{natbib}
\usepackage{comment}
\usepackage{caption}
\usepackage{subcaption}
\usepackage{algorithm}
\usepackage{algpseudocode}
\usepackage{physics}
\usepackage[autostyle]{csquotes}
\usepackage{dirtytalk}
\usepackage{lineno}

\newcommand{\mbs}[1]{\boldsymbol{#1}}

\def\bA{{\mbs{A}}}  \def\bC{{\mbs{C}}}
 \def\bE{{\mbs{E}}} \def\bF{{\mbs{F}}}
\def\bG{{\mbs{G}}}  \def\bI{{\mbs{I}}}

\def\bP{{\mbs{P}}}  
  
 \def\bX{{\mbs{X}}} 
 \def\b0{{\mbs{0}}}
\def\ba{{\mbs{a}}}

 \def\bn{{\mbs{n}}} 
  \def\br{{\mbs{r}}}
  \def\bu{{\mbs{u}}}
 \def\bx{{\mbs{x}}}

\def \AE {{A^{\bE}}}

\def	\sigm	{{\sigma}}

\def    \tr {\textrm{tr\,}}

\begin{document}

\begin{frontmatter}

\title{Digital twin model of colon electromechanics for manometry prediction of laser tissue soldering}

\author[inst1]{René Thierry Djoumessi}
\author[inst1]{Pietro Lenarda\corref{cor1}}
\author[inst2]{Alessio Gizzi}
\author[inst3]{Simone Giusti}
\author[inst4]{Pietro Alduini}
\author[inst1]{Marco Paggi}

\affiliation[inst1]{organization={IMT School for Advanced Studies Lucca},
            addressline={Piazza San Francesco 19}, 
            city={Lucca},
            postcode={55100}, 
            country={Italy}}

\affiliation[inst2]{organization={Department of Engineering, Universita' Campus Bio-Medico di Roma},
            addressline={Via A. del Portillo 21}, 
            city={Rome},
            postcode={00128}, 
            country={Italy}}
            
\affiliation[inst3]{organization={Sigma Ingegneria s.r.l.},
            addressline={Via di Sant'Alessio 1957}, 
            city={Lucca},
            postcode={55100}, 
            country={Italy}}     

\affiliation[inst4]{organization={San Luca Hospital},
            addressline={Via Guglielmo Lippi Francesconi, 556}, 
            city={Lucca},
            postcode={55100}, 
            country={Italy}}

\cortext[cor1]{Corresponding author Email addresses: pietro.lenarda@imtlucca.it (P. Lenarda), rene.djoumessi@imtlucca.it (R. T. Djoumessi)}

\begin{abstract}
The present study introduces an advanced multi-physics and multi-scale modeling approach to investigate in silico colon motility. We introduce a generalized electromechanical framework, integrating cellular electrophysiology and smooth muscle contractility, thus advancing a first-of-its-kind computational model of laser tissue soldering after incision resection. The proposed theoretical framework comprises three main elements: a microstructural material model describing intestine wall geometry and composition of reinforcing fibers, with four fiber families, two active-conductive and two passive; an electrophysiological model describing the propagation of slow waves, based on a fully-coupled nonlinear phenomenological approach; and a thermodynamical consistent mechanical model describing the hyperelastic energetic contributions ruling tissue equilibrium under diverse loading conditions. The active strain approach was adopted to describe tissue electromechanics by exploiting the multiplicative decomposition of the deformation gradient for each active fiber family and solving the governing equations via a staggered finite element scheme. The computational framework was fine-tuned according to state-of-the-art experimental evidence, and extensive numerical analyses allowed us to compare manometric traces computed via numerical simulations with those obtained clinically in human patients. The model proved capable of reproducing both qualitatively and quantitatively high or low-amplitude propagation contractions. Colon motility after laser tissue soldering demonstrates that material properties and couplings of the deposited tissue are critical to reproducing a physiological muscular contraction, thus restoring a proper peristaltic activity. 
\end{abstract}


\end{frontmatter}


\section{Introduction}
\label{sec:Intro1}
Anastomotic leakage following resection of gastrointestinal (GI) lesions is a primary source of concern for clinicians and patients alike \citep{turrentine2015morbidity}. Nowadays, developing techniques to improve tissue resection and sealing, thus reducing leakage rates, are essential to increase the reliability of interventions in a clinical setting. Current methods of tissue fixation, such as sutures or staples, exert tensile and compressive forces on the attached tissue, potentially forming gaps in the anastomotic line, resulting in anastomotic leakage and/or pathological scars \citep{gizzi2012,chadi2016emerging,pierfranco2018}. In recent years, laser technology has been used as an alternative method for tissue bonding in different tissues such as skin, cornea, buccal mucosa, and even nerves using laser tissue soldering (LTS) and laser tissue welding (LTW) \citep{gerasimenko2022reconstruction,birkelbach2020vitro,basov2019strong} principally. The advantage of these techniques is that they allow to glue tissues that have been bio-printed \citep{chirianni2024}. In this regard, patches can be 3D printed externally, or, thanks to new emerging technologies protected by patents \citep{Alduini2020}, they can directly bio-printed within the organs thanks to a new generation of endoscopes. 

However, bonding requires great dexterity on the part of the clinician and the robots, as parameters such as compressive force, welding material, and laser temperature must be taken into account for the operation to be successful \citep{nisky2015teleoperated, sharon2018expertise}. Moreover, after such an operation, there is always a risk of local inflammation and scarring of the surrounding tissues \citep{huang2013laser}. This inflammatory response can lead to muscle fibrosis in the colon, resulting in increased muscle stiffness surrounding the welding tissue. On the other hand, it can happen that after LTS, the weld area is not effective enough (not fully cellularized and/or integrated) to withstand the burst pressure or that the weld eventually ruptures at very low pressure \citep{ashbell2023laser,urie2015gold,huang2013laser,mushaben2018spatiotemporal}. Such a condition generally happens when the compressive force during LTS, the soldering time, and the laser power are not properly designed for the patient.

Therefore, there is an urgent need to develop a high-fidelity digital twin model of the colon that can be exploited to simulate and predict the response of the organ after such interventions, understand how to optimize the 3D bio-printed patch and identify potential critical issues. A digital twin model can effectively enable digital technology to accelerate the move from the technological proof of concept to the clinical setting.

In order to provide data of clinical use, any digital twin model designed for such a purpose should be able to accurately predict the effect of tissue resection and repairing on High-Resolution Manometry (HRM).   
HRM is a standard GI motility diagnostic system that measures intraluminal pressure activity in the GI tract using a series of closely spaced pressure sensors \citep{conklin2009color, dinning2015use, li2019high}. Displayed and interpreted by intraluminal pressure topography (EPT) spatiotemporal patterns, HRM/EPT provides a detailed assessment of GI function that is critical in evaluating patients with nonobstructive dysphagia. Accordingly, esophageal motility diagnoses are determined systematically by applying objective metrics of peristaltic function to the Chicago Classification of Motility Disorders \citep{yadlapati2021esophageal,noh2023comparison}. In general, intestinal dysmotility is characterized by altered motility patterns that result in compromised transit of luminal contents accounting for 30-45\% of gastrointestinal conditions globally \citep{rai2021prokinetics}. It is worth noticing that, in clinical practice, HRM is the primary method used to evaluate GI motor function invasively. In the case of colon assessment, HRM involves inserting a catheter with 36 pressure transducers spaced $1 \; \rm cm$ or $2 \; \rm cm$ apart, sometimes for hours.

Regarding the underlying biophysics mechanisms, colon contraction is ruled by electrophysiological slow waves generated by the coordination between interstitial cells of Cajal (ICCs) and smooth muscle cells (SMCs) \citep{huizinga2009gut,sanders2016regulation, sanders2006interstitial, tremain2024endoscopic}. Several electrical models have been proposed in the literature to reproduce the complex spatiotemporal phenomenology of gastrointestinal excitation \citep{buist2010extended,du2010multiscale,du2013model,lees2011biophysically,corrias2013modelling,athavale2023computational,athavale2024neural}. Moreover, the mechanical activity of the gastrointestinal system is ensured by the interaction between ICCs and SMCs from the cellular to the organ level \citep{corrias2007quantitative, corrias2008quantitative, yeoh2017modelling,aliev2000simple,brandstaeter2018computational, gizzi2010electrical}. The electric waves produced by ICCs propagate to the surrounding SMCs via dedicated gap junctions proteic nanostructures \citep{hanani2005intercellular}. These localized plasma membrane fusions provide direct electrical cell-cell coupling, forming what is known as a functional syncytium. SMC contraction occurs when neuronal/hormonal signals coincide with slow wave electrical phases. In particular, voltage membrane depolarisation activates L-type voltage-gated calcium channels \citep{sanders2016regulation}, which is the initial event triggering complex mechanisms on several scales: the opening of the Calcium channels triggers the entry of Calcium ions, leading to the contraction of the smooth muscle that deforms the GI wall. 

\begin{figure}[]
    \centering
    \includegraphics[scale=0.5]{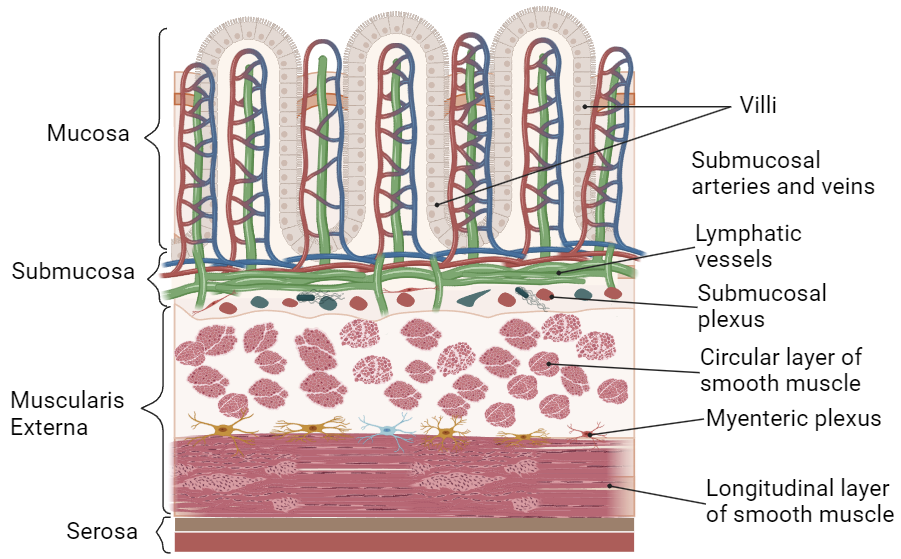}
    \caption{Structure of the Gastrointestinal wall highlighting the different layers with their internal microstructure.}
    \label{fig:0}
\end{figure}

The GI wall is a complex multilayered structure (see Fig. \ref{fig:0}) comprising: 
serosa (outermost layer), a simple epithelium secreting serous fluid;
muscularis externa, containing longitudinal and circular SMC fibers;
Auerbach’s plexus containing enteric neurons;
submucosa, a dense layer of connective tissues containing large blood and lymphatic vessels;
mucosa, formed by three sublayers (epithelium, lamina propria, and muscularis mucosae) and containing villi and microvilli to maximise the exchange surface.
Accordingly, various constitutive models have been proposed to reproduce the electromechanical behavior of the different GI sections, e.g., stomach \citep{brandstaeter2018computational, papenkort2023geometry,toniolo2022coupled,friis2023biomechanical}, small intestine \citep{du2013model, gizzi2010electrical,hosseini2020biomechanics, aydin2017experimental}, and colon \citep{pandolfi2017visco,johnson2019much}. Though active electromechanics has been proposed in few cases, most of the literature is based on the active stress approach disregarding multiple anisotropic components and lacking a robust numerical implementation.

We assume the constitutive model for the passive part as an exponential Holzapfel-type material models described in \citep{patel2022biomechanical, puertolas2020comparative} since these structure-based approaches account for directional fiber reinforcements. Overall, exponential anisotropic constitutive laws have been shown to characterize well the mechanical behavior of several intact GI segments (esophagus, small intestine, large intestine, and rectum), validating their performances against uniaxial (planar uniaxial extension, planar shear) and biaxial tests (planar biaxial extension and tubular inflation-extension).

To the best of the authors' knowledge, a comprehensive computational study investigating the mechanical effects of the LTS procedure on colonic motility and dysmotility, addressing the resulting manometry patterns, has not yet been proposed in the literature. To fill this gap, we advance detailed mathematical modeling of colon motility. 

In the present work, we propose an active strain electromechanical model of colon motility considering an anisotropic hyperelastic constitutive law, consisting of four reinforcing microstructures, thus embedding two active muscle fibers (longitudinal and circumferential) and the two passive collagen sheets present in the submucosa layer, coupled with a phenomenological electrophysiological model that finely reproduces SMCs and ICCs spatiotemporal dynamics. The proposed multi-field model is employed to numerically study the effect of deposited bio-printed material or albumin due to LTS endoscopic resection represented by an elliptical portion on the inner part of the colon wall and characterized by altered material properties. In particular, we show that the computational framework is able to reproduce the intraluminal pressure maps corresponding to HRM data, both health and disease. 

The manuscript is organized as follows. 
In section \ref{sec:model}, the active strain electro-mechanical formulation for colon motility is presented. 
In section \ref{strongnumeric}, the strong form of the problem is derived, and the finite element discretization and associated staggered solver are presented.
In section \ref{sec:simu}, numerical analyses are carried out, as well as the exploitation of the digital twin model to investigate the effect of laser tissue soldering on colon motility.
Conclusions, limitations, and perspectives are discussed in section \ref{sec:con}.

\section{Electromechanical constitutive modeling of colon motility}
\label{sec:model}
In this section, we recall the governing equations for active strain finite deformations coupled with GI electrophysiology. 

We represent a scalar, a vector, and a second-order tensor with the lowercase letters ($a$), lowercase bold letters ($\ba$) and capital bold letters ($\bA$), respectively, and ($\bA^T$) stands for the transpose of a tensor. According to the tensor notation, we indicate the scalar product with $(\cdot)$, the double contraction with $(:)$, and the dyadic product with $(\otimes)$. Moreover, $\nabla$, $\nabla\cdot$ and $\nabla^2$ represent the gradient, divergence, and Laplace operator, respectively.

\subsection{Finite kinematics}
The kinematics of the deformable GI tissue is embedded in the classical description of continuum mechanics under the assumption of finite elasticity. Let $\bX$ denotes the material position vector in the reference (undeformed) configuration $\Omega_0\subset\mathbb{R}^d$, $ d=2,3 $ at time $t=0$, and $\bx=\bX+\bu$ stands for the spatial position vector in the current (deformed) configuration  $\Omega_t\subset\mathbb{R}^d$,$ d = 2,3$ at time $t \in [0, T]$, whereas $\bu$ is the displacement field. Hereby, the material point $\bX$ is linked to the spatial point $\bx$ in the current configuration by a nonlinear deformation map $\varphi_t: \Omega_0 \rightarrow \Omega_t$, $\bX \rightarrow \bx$. The deformation gradient associated with $\varphi_t$ is defined as $\bF={\partial\bx}/{\partial\bX}$ and its Jacobian is $J=\det\bF > 0$. The left Cauchy-Green deformation tensor is defined as $\bC=\bF^T \bF$. The first isotropic invariant of deformation is defined as $I_1(\bC)=\tr(\bC)$, where $\tr( \cdot )$ denotes the trace operator, and the anisotropic fourth pseudo-invariant reads $I_4(\bC)=\bC:\bG$, with $\bG$ the structure tensor.

The contraction of an excitable biological tissue combines active and passive behaviours, nonlinearly coupling electrophysiological cellular dynamics with a material hyperelastic response. According to the active strain approach \citep{brandstaeter2018computational, cherubini2008electromechanical, ambrosi2011electromechanical,ruiz2020thermo}, a multiplicative decomposition of the deformation gradient tensor into an elastic, $\bF_e$, and an inelastic, $\bF_a$, part is put forward:
\begin{equation} \label{eq:1}
    \bF = \bF_e \bF_a \,.
\end{equation}
Such an approach allows to apply multiscale and multiphysics couplings in a homogenised continuum framework over the local active deformation map.

Stemming from the gastric microstructural approach detailed in \citep{brandstaeter2018computational}, and according to the active strain kinematics, we consider longitudinal and circumferential SMC directions as contractile units ruled by the active part of the deformation gradient:
\begin{equation} \label{eq:2}
    \bF_a = \bI-\gamma(V)(\alpha_c \bn_c \otimes  \bn_c + \alpha_l \bn_l \otimes \bn_l ) + \gamma_n \bn_n \otimes \bn_n \,,
\end{equation}
where $\bn_c$, and $\bn_l$ are the orthonormal unit vectors in the circumferential and longitudinal direction, respectively, while $\bn_n = \bn_c \times \bn_l$ represents the unit vector orthogonal to their plane. In Eq.~\eqref{eq:2}, $\alpha_c$ and $\alpha_l$ stand for material parameters ruling the amount of contraction in a certain direction, while $\gamma_n$ enforces the incompressibility constraint in such a way that $\det (\bF_a) =1$, i.e.:
\begin{equation} \label{eq:3}
   \gamma_n = \frac{1-(1-\gamma\alpha_c)(1-\gamma\alpha_l)}{(1-\gamma\alpha_c)(1-\gamma\alpha_l)} \,.
\end{equation}

The excitation function $\gamma(V)$ couples the mechanical problem with the electrophysiological one via a smooth activation function, defined in \citep{patel2022biomechanical}, dependent on the active membrane potential $V$ crossing the smooth muscle layer: 
\begin{equation}\label{eq:4}
    \gamma(V) = (1-e^{1-\beta_1(V-V_{th})})(1-e^{1-\beta_2(V-V_{th})}) H(V-V_{th}),
\end{equation}
where, $\beta_1$, $\beta_2$, and $V_{th}$ are the material parameters linked to the intracellular Ca$^{2+}$ dynamics, while $H(V-V_{th})$ is a Heaviside step function switching on active contraction whenever the threshold $V_{th}$ is reached. The parameters of the function $\gamma(V)$ and inelastic tensor can be found in the following Table \ref{tab:active}. 
\begin{table}[h!]
    \centering
    \caption{Material parameters of the active strain model.}
    \begin{tabular}{c c c c c}
    \hline
     $\alpha_c$  & $\alpha_l$ & $\beta_1$ & $\beta_2$ & $V_{th}$\\
     \hline
        $0.2$ & $0.2$ & $10$ & $10$ & $50\% V$\\
    \hline
    \end{tabular}
    \label{tab:active}
\end{table}

\subsection{GI electrophysiological model}

The electrophysiological model adopted in \citep{gizzi2010electrical} is herein recalled and generalized. The SMC and ICC layers are labeled with indices $s$ and $i$, respectively. The resulting system of nonlinear partial differential reaction-diffusion equations describe the coupled dynamics between the transmembrane potential variables, $u_s, u_i$, and the slow currents ones, $v_s, v_i$:
\begin{linenomath}
\begin{subequations}
\begin{align}
\frac{\partial u_s}{\partial t} &= f(u_s)+D_s\nabla^2 u_s-v_s + F_s(u_s,u_i) + I_{stim}^s \quad \textrm{on} \quad \Omega_0 \times [0,T], \label{eq:14a}
\\
\frac{\partial v_s}{\partial t}  &= \epsilon_s[\lambda_s (u_s-\beta_s)-v_s] \quad \textrm{on} \quad \Omega_0 \times [0,T], \\
\frac{\partial u_i}{\partial t} &= g(u_i)+D_i\nabla^2 u_i-v_i + F_i(u_s,u_i) + I_{stim}^i \quad \textrm{on} \quad \Omega_0 \times [0,T], \label{eq:14c}\\
\frac{\partial v_i}{\partial t}  &= \epsilon_i(z)[\lambda_i (u_i-\beta_i)-v_i] \quad \textrm{on} \quad \Omega_0 \times [0,T],
\end{align}
\label{eq:5}
\end{subequations}
\end{linenomath}
where:
\begin{linenomath}
\begin{subequations}
\begin{align}
f(u_s) &=  k_su_s(u_s-a_s)(1-u_s) \,, \quad
& F_s(u_s,u_i) = \alpha_s D_{si}(u_s-u_i) \,, \\
g(u_i) &=  k_iu_i(u_i-a_i)(1-u_i) \,, \quad
& F_i(u_s,u_i) = \alpha_i D_{is}(u_s-u_i) \,.
\end{align}
\label{eq:6}
\end{subequations}
\end{linenomath}
Here, $I_{stim}^s$ and $I_{stim}^i$ are the stimulation currents applied to the SMC and ICC respectively; $D_s, D_i$ are the diffusivities (assumed isotropic); $\lambda_s, \lambda_i$ are the coupling factors between the membrane potential and recovery variable; $D_{si}, D_{is}$ are the diffusivities of the gap junctions between the two cell species; $k_i, k_s, a_s, a_i,\alpha_s, \alpha_i$ are phenomenological model parameters and their values are provided in Table.~\ref{table:1}.
The parameter $\epsilon(z)$, which is proportional to the oscillation frequency of the ICCs cells, represents a space-dependent excitability function, decreasing with distance from the pylorus, in agreement with the experimental interpolation plot reported in \citep{aliev2000simple}. 

\begin{table}[h!]
\centering
\caption{Electrophysiological parameters adapted from \citep{aliev2000simple,gizzi2010electrical}.}
\begin{tabular}{c c c c} 
 \hline
\multicolumn{2}{c}{SMC layer} &\multicolumn{2}{c}{ICC layer} \\
 \hline
 $k_s$=$10$ & $a_s$=$0.06$ &$k_i$=$7$ & $a_i$=$0.5$ \\ 
 $\beta_s$ = $0$ & $\lambda_s$=$8$ & $\beta_s$ = $0.5$ & $\lambda_i$=$8$ \\
 $\epsilon_s$=$0.15$ & $\alpha_s$=$1$ & $\epsilon_i$=$\epsilon_i(z)$ & $\alpha_i$=$-1$ \\
 $D_{si}$=$0.3$ & $D_s$=$0.4$ & $D_{is}$=$0.3$ & $D_i$=$0.04$ \\
 \hline
\end{tabular}
\label{table:1}
\end{table}

In view of coupling the electrophysiological model with the active deformation map via $\gamma(V)$ in Eq.~\eqref{eq:4}, the action potential $V$ is identified with the transmembrane voltage dynamics $u_s$ passing through the SMC layer.

\subsection{Active mechanics of the GI system}
Considering a structure-based constitutive formulation, we proceed with an additive decomposition of the elastic strain energy density into isotropic and anisotropic parts:
\begin{equation}\label{eq:7}
    \Psi = \Psi^{ \rm iso} + \Psi^{ \rm aniso} \,,
\end{equation}
where, the isotropic contribution is related to the passive mechanical response of the non-collagenous components of the tissue (matrix) \citep{puertolas2020comparative,sokolis2021variation}. For the sake of clarity, and with no loss of generality (other options could be made with similar results \citep{nagaraja2021phase}), in the following we restrict our analysis to the Neo-Hookean material model characterized by the following isotropic strain energy density function:
\begin{equation} \label{eq:8}
    \Psi^{\rm iso} = \mu(I_1 -3) \,,
\end{equation}
with $\mu$ the passive isotropic stiffness. 

\begin{figure}[h]
    \centering
    \includegraphics[scale=0.55]{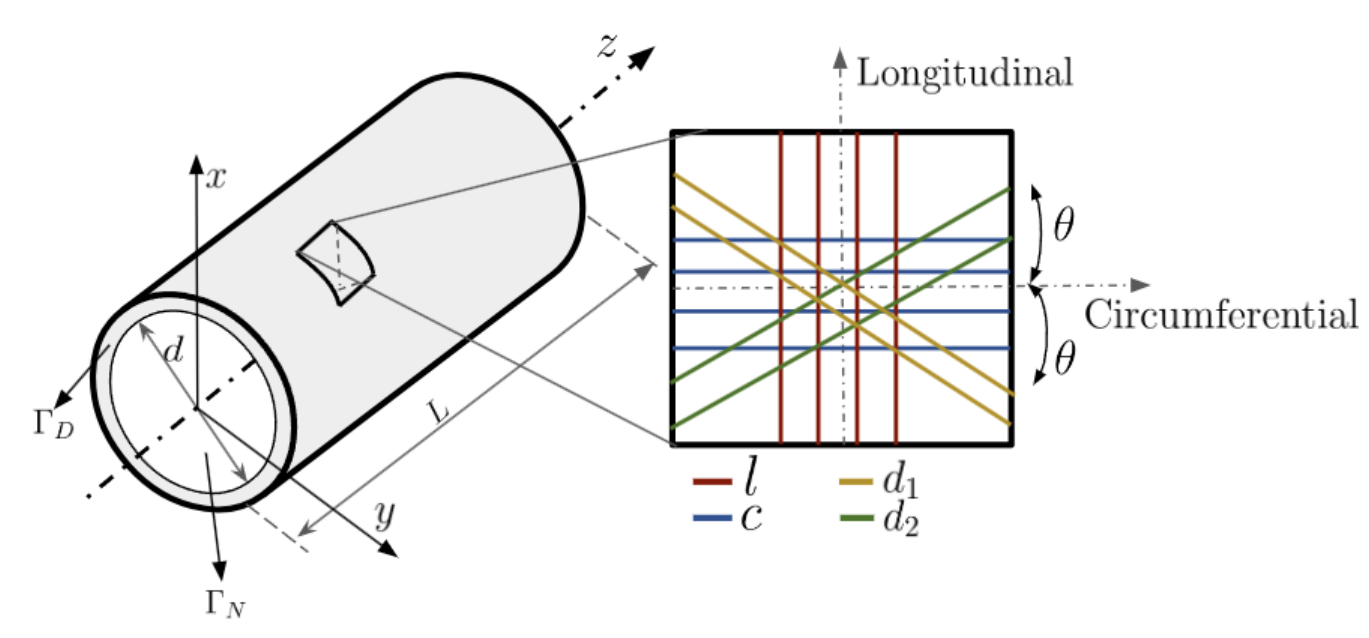}
    \caption{Idealized colon segment with length $L$ and diameter $d$. The zoomed cross-section represents the wall microstructure, which is composed of four families of fibers embedded in an isotropic elastin matrix. The directions of the fibers are uniquely defined with respect to the circumferential direction by the angle $\theta$; $l$ represents the external longitudinal muscular layer, $c$ the internal circumferential muscular fiber, $d_1$ and $d_2$ are the submucosa helically collagen fibers.}
    \label{fig:1}
\end{figure}

The anisotropic energetic component presents, in general, passive and active contributions. The passive part is associated with the mechanical response of directional collagen fibers. These are assumed to mimic the submucosa reinforcement, $d_1$ and $d_2$, oriented with an angle $\theta$ according to the circumferential directions (see Fig.~\ref{fig:1}). The active anisotropic contribution is due to the presence of SMC fibers in the longitudinal, $l$, and circumferential, $c$, direction, respectively:
\begin{linenomath}
\begin{align*} \label{eq:9}
  \Psi^{\rm aniso} 
  &= 
  \Psi^{\rm aniso}_{\rm p} +
  \Psi^{\rm aniso}_{\rm act}  
  \\
  &= 
  \sum_{i \in \{ d_1, d_2 \}} \frac{k_1^i}{4k_2^i}[e^
  {k_2^i(I_4^i -1)_+^2}-1] +
  \sum_{i \in \{ l,c\}} \frac{k_1^i}{4k_2^i}[e^
  {k_2^i(I_4^i -1)_+^2}-1]  
  .
\end{align*}
\end{linenomath}
Here, the notation $(y)_+:= y$ if $y\geq 0$ or near to zero otherwise for any generic real value of $y$ reproduces the tension-compression switch approximation, considering the anisotropic contribution negligible in compression \citep{patel2022biomechanical}. The anisotropic fourth invariant of the deformation $I_4^j$ measures the stretch for each fiber family $j$ along their main direction, as:
\begin{equation} \label{eq:9}
    I_4^j = \bC:(\bn_j \otimes \bn_j) 
    \,, \quad j \in \{l,c,d_1,d_2 \},
\end{equation}
where $\bn_j$ are the unit normal vectors associated to each fiber direction $j$.

The material parameters  $k_1^j$ (stiffness-like) and $k_2^j$ (nondimensional) are associated with the directional behavior of the material. According to previous studies \cite{puertolas2020comparative,sokolis2021variation, nagaraja2021phase}, the material parameters for the diagonal fibers are assumed identical. Furthermore, a preliminary tuning analysis was conducted to identify experimental-based material stiffness as explained in \ref{sec:B}. Material parameters are provided in Table \ref{table:T2}.

According to the given prescriptions, the first Piola-Kirchhoff stress tensor is derived under variational priciples as:
\begin{equation} \label{eq:10}
    \bP 
    = 
    \frac{\partial \Psi}{\partial \bF} -pJ\bF^{-T} 
    = 
    {\rm det} (\bF_a) \left(\frac{\partial \Psi^{ \rm iso}}{\partial \bF_e}+ \frac{\partial \Psi^{ \rm aniso}}{\partial \bF_e}\right) \bF_a^{-T}
    -
    pJ\bF^{-T} \,,
\end{equation}
where $p$ stands for the solid hydrostatic pressure.

It is worth mentioning that from a numerical viewpoint, the precise creation of the fibrous structure of the GI tract is crucial. Due to their complex structure and strong layer adhesion, a homogenized layer was chosen for fiber generation throughout the thickness of the wall. Similarly to the method presented in \citet{piersanti2021modeling}, we adopted and customized a rule-based algorithm \citep{rossi2014thermodynamically}, originally proposed for cardiac fibers, to reproduce colon microstructure and generate muscle fibers throughout all the simulations (detailed explanation is provided in \ref{sec:C}).

\section{Numerical implementation}
\label{strongnumeric}
\subsection{Strong form of the problem}

The strong form of the problem is given by the following set of nonlinear coupled partial differential equations prescribing mechanical equilibrium in terms of displacement field $\bf u$ and pressure $p$ variables, defined in the undeformed colon domain $\Omega_0$:
\begin{linenomath}
\begin{subequations}
\begin{align}
    \nabla \cdot \bP &= {\bf 0} \,,  
    \quad \textrm{on} \quad \Omega_0 \times [0,T] \label{eqn:line-1} 
    \\
    J-1 &=0 \,, \quad 
    \textrm{on} \quad \Omega_0 \times [0,T] \label{eqn:line-2} 
\end{align}
\label{eq:11}
\end{subequations}
\end{linenomath}
complemented with electrophysiological balance laws Eq.~\eqref{eq:5}, solved for the electrophysiological variables $u_s, v_s, u_i, v_i$.

Mixed boundary conditions of normal displacement and traction close the system:
\begin{linenomath}
\begin{subequations}
\begin{align}
    \bu \cdot \bn &=  0 \,, 
    \quad \text{on} \quad \Gamma_{D} \times [0,T]
    \label{eq:bc1}
    \\
    \bP \bn -p_0J\bF^{-T} \bn &= {\bf 0} \,,
    \quad \text{on} \quad \Gamma_{N} \times [0,T]
    \label{eq:bc2}
\end{align}
\label{eq:12}
\end{subequations}
\end{linenomath}
being $\Gamma_{D}$ and $\Gamma_{N}$ a disjoint partition of the boundary. Condition Eq.~\eqref{eq:bc1} constraints normal motion along the surface $\Gamma_{D}$. The term $p_0$ in Eq.~\eqref{eq:bc2} denotes a prescribed (possibly time-dependent) boundary load (normal stress--pressure) associated with the presence of digesta within the lumen. In the present case, such a load is assumed uniform over the deformed counterpart of $\Gamma_{N}$, and applied in the normal direction to the internal surface of the colon in the deformed configuration.

Finally, no flux boundary conditions are considered for the electrophysiological problem:
\begin{linenomath}
\begin{equation}
D_s\nabla u_s \cdot \bn= 0, \quad D_i\nabla u_i \cdot \bn=  0 \quad \textrm{on} \quad \partial\Omega_N,
\label{eq:13}
\end{equation}
\end{linenomath}
where $\partial\Omega_N$ stands for the Neumann boundary (the whole boundary for the electrophysiological problem).

\subsection{Mixed‑primal weak variational form}
\label{sec:FEM} 
The trial spaces, where the solution of the weak form of the problem is defined, are given by:
\begin{linenomath}
\begin{equation}
\bu \in \mathbf{V}:=L^2(0,T;\mathbf{H}^1(\Omega_0)), \; p \in Q:= L^2(0,T; L^2(\Omega_0)), 
\label{eq:14}
\end{equation}
\end{linenomath}
for displacements and pressure and
\begin{linenomath}
\begin{equation}
(u_s,u_i,v_s,v_i) \in V^4 :=[ L^2(0,T; H^1(\Omega_0)) ]^4,
\label{eq:15}
\end{equation}
\end{linenomath}
for the electrophysiological variables. 
The virtual displacements for the mechanical problem are introduced as $ \delta \bu \in \mathbf{V}_0$, as well as the test functions for the pressure $ \delta p \in Q_0 $ and $(\delta u_s, \delta u_i, \delta v_s, \delta v_i) \in (V_0)^4$ for the electrophysiological problem, defined on the spaces of the corresponding fields, and vanishing on the Dirichlet part of the boundary.
Multiplying Eq.~\eqref{eq:11} by a virtual displacement, using the divergence theorem and the boundary conditions in Eq.~\eqref{eq:12}, the weak form of the mechanical problem is: \\
find displacement $\bu \in \mathbf{V}$ and pressure $p \in Q$ such that
\begin{linenomath}
\begin{subequations}
\begin{align}
\int_{\Omega_{0}}\bP:\nabla \delta \bu -\int_{\Gamma_{N}} p_{0}(t)J\bF^{-T} \bn \cdot \delta \bu=0 ,\ \forall \delta \bu \in \mathbf{V}_0 \,,\\
\int_{\Omega_{0}}(J-1)\delta p + \int_{\Omega_{0}}\zeta_{stab}\nabla p \cdot \nabla \delta p=0 ,\ \forall \delta p \in Q_0 \,,
\end{align}
\label{eq:16}
\end{subequations}
\end{linenomath}
\noindent where $\zeta_{stab}$ is a positive pressure stabilization parameter used as a locking-free parameter to enhance the stability of the discrete problem \cite{chavan2007locking}. 
The mechanical problem can be rewritten in a more compact form as: 
\\
find  displacement and pressure $\bu$ and $p$ such that
\begin{linenomath}
\begin{multline}
    \mathcal{M}(\bu,p;\delta \bu,\delta p) := \int_{\Omega_{0}}\bP:\nabla\delta \bu  -\int_{\Gamma_{N}} p_{0}(t)J\bF^{-T} \bn \cdot \delta \bu +
    \\
    \int_{\Omega_{0}}(J-1)\delta p 
    + \int_{\Omega_{0}}\zeta_{stab}\nabla p \cdot \nabla \delta p=0 \,,
\label{eq:17}
\end{multline}
\end{linenomath}
for all test functions $\delta \bu$ and $\delta p$.
Analogously, multiplying the rest of Eq.~\eqref{eq:11} by test functions $(\delta u_s, \delta v_s,\delta u_i,\delta v_i) \in (V_0)^4$, applying the divergence theorem and the condition of zero flux at the boundary in Eq.~\eqref{eq:13}, the weak form of the electrophysiology problem can be written as it follows: at each time step $t^{n+1}=t^n +\Delta t$ of an equispaced partition of the time interval $[0, T]$, given the solution of the electrophysiology problem at the previous timestep $(u^{n}_s,v^{n}_s,u^{n}_i,v^{n}_i) \in V^4$ find the vector $(u^{n+1}_s,v^{n+1}_s,u^{n+1}_i,v^{n+1}_i) \in V^4$ at the current timestep $t^{n+1}$ such that it is satisfied
\begin{linenomath}
\begin{subequations}
\begin{align}
\int_{\Omega_{0}}\frac{u_s^{n+1}-u_s^{n}}{\Delta t}\delta u_s+\int_{\Omega_{0}}D_s\nabla u_s^{n+1}\cdot\nabla\delta u_s=\int_{\Omega_{0}}I_{ion}^s (u_s^{n},v_s^{n},u_i^{n})\delta u_s \,,\\
\int_{\Omega_{0}}\frac{v_s^{n+1}-v_s^{n}}{\Delta t}\delta v_s =\int_{\Omega_{0}}R_{s}(u_s^{n},v_s^{n})\delta v_s \,, \\
\int_{\Omega_{0}}\frac{u_i^{n+1}-u_i^{n}}{\Delta t}\delta u_i+\int_{\Omega_{0}}D_i\nabla u_i^{n+1}\cdot\nabla\delta u_i=\int_{\Omega_{0}}I_{ion}^i (u_s^{n},v_i^{n},u_i^{n})\delta u_i \,,\\
\int_{\Omega_{0}}\frac{v_i^{n+1}-v_i^{n}}{\Delta t}\delta v_i =\int_{\Omega_{0}}R_{i}(u_i^{n},v_i^{n})\delta v_i \,,
\end{align}
\label{eq:18}
\end{subequations}
\end{linenomath}
for all test functions $(\delta u_s, \delta v_s,\delta u_i,\delta v_i) \in (V_0)^4$, where: 
\begin{linenomath}
\begin{subequations}
\begin{align}
I_{ion}^i (u_s,v_i,u_i) &=  g(u_i)-v_i + F_i(u_s,u_i) + I_{stim}^i \,,\\
R_{i} (u_i,v_i) &= \epsilon_i(z)[\lambda_i (u_i-\beta_i)-v_i] \,, \\
I_{ion}^s (u_s,v_s,u_i) &=  f(u_s)-v_i + F_s(u_s,u_i) + I_{stim}^s \,,\\
R_{s} (u_s,v_s) &= \epsilon_s[\lambda_i (u_s-\beta_s)-v_s] \,.
\end{align}
\label{eq:19}
\end{subequations}
\end{linenomath}
The implicit Euler scheme for the discretization of the time derivative has been adopted in Eqs.~\eqref{eq:18}, while an explicit treatment of the reaction terms has been used. 
In a more compact notation, the electrophysiology problem can be written as: find $u_s$, $u_i$, $v_s$ and $v_i$ such that
\begin{linenomath}
\begin{equation}
\label{eq:E}
    \mathcal{E}(u_s,u_i,v_s,v_i;\delta u_s, \delta u_i, \delta v_s, \delta v_i):= \mathcal{E}_1+ \mathcal{E}_2+ \mathcal{E}_3+ \mathcal{E}_4=0 \,,
\end{equation}
\end{linenomath}
for all test functions $\delta u_s, \delta u_i, \delta v_s, \delta v_i$,
where: 
\begin{linenomath}
\begin{align}
\mathcal{E}_1 &= \mathcal{E}_1(u_s,u_i,v_s;\delta u_s, \delta u_i, \delta v_s) \,, \\
\mathcal{E}_2 &=\mathcal{E}_2(u_s,v_s;\delta u_s,  \delta v_s) \,, \\
\mathcal{E}_3 &= \mathcal{E}_3(u_s,u_i,v_i;\delta u_s, \delta u_i, \delta v_i) \,, 
\\\mathcal{E}_4 &= \mathcal{E}_4(u_i,v_i;\delta u_i,  \delta v_i) \,,
\end{align}
\end{linenomath}
are respectively the residuals of Eqs.~\eqref{eq:18}.

\subsection{Finite element discretization}
\label{sec:Numer1}

The computational domain $\Omega_0$ has been discretized into tetrahedral finite elements and the unknowns have been approximated using Lagrangian shape functions $\mathbb{P}_2$ and $\mathbb{P}_1$ for the two variational problems defined by Eqs.~\eqref{eq:17} and \eqref{eq:E}, respectively. The problem has been implemented in the open-source finite element software \texttt{FEniCS} \citep{alnaes2015fenics}.
A splitting scheme was adopted to solve separately the mechanical problem Eq.~\eqref{eq:17} and the electrophysiology problem Eq.~\eqref{eq:E}, the nonlinear mechanical problem is solved using the Newton-Raphson method and, at each Newton's iteration, the resulting linear system given by
$d \mathcal{M}(\Delta \bu, \Delta p;  \delta \bu,\delta p)=-\mathcal{M}(\bu_k^{n+1},p_k^{n+1};\delta \bu,\delta p)$ for the corrections $\Delta \bu$ and $\Delta p$ is solved using a \texttt{BiCGStab} (Biconjugate Gradient Stabilised) method preconditioned with \texttt{ILU} (Incomplete LU factorization). The tangent operator $d \mathcal{M}$ associated to the nonlinear variational mechanical problem in Eq.~\eqref{eq:17} has been computed via the symbolic derivative \texttt{derivative}.
 For the solution of electrophysiological problem, the \texttt{PETSc} library (Portable, Extensible Toolkit for Scientific Computing) has been used.
The algorithm for the solution of the coupled electro-mechanical problem describing the colonic motility is detailed in Alg.~\ref{alg:cap}.

\begin{algorithm}
\caption{Algorithm for the electro-mechanical motility of a GI tract}\label{alg:cap}
\begin{algorithmic}[1]
\State  \textbf{Input} Initial and boundary conditions for displacement, pressure and electrophysiological variables:
\While{$t^n < T$}
     \State \textbf{Given:} displacement and pressure $\bu^n, p^n$ solve the linear electrophysiology problem: $\mathcal{E}(u^{n+1}_s,u^{n+1}_i,v^{n+1}_s,v^{n+1}_i;\delta u_s, \delta u_i, \delta v_s, \delta v_i)=0$ 
      \State \textbf{Update} EP solutions $(u^{n}_s,u^{n}_i,v^{n}_s,v^{n}_i) \gets (u^{n+1}_s,u^{n+1}_i,v^{n+1}_s,v^{n+1}_i)$
     \State \textbf{Given:} the electrophysiological variables $(u^{n+1}_s,u^{n+1}_i,v^{n+1}_s,v^{n+1}_i)$, solve the mechanical problem via Newton-Raphson procedure:
     \For {a given Newton iteration $k$ }
           \State \textbf{Given:} $\bu_k^{n+1}$ and $p_k^{n+1}$ solve the linearized mechanical problem: $d \mathcal{M}(\Delta \bu, \Delta p;  \delta \bu,\delta p)=-\mathcal{M}(\bu_k^{n+1},p_k^{n+1};\delta \bu,\delta p)$ 
           \State \textbf{Set}: $\bu^{n+1}_{k+1}=\Delta \bu+\bu^{n+1}_{k}$ and $p^{n+1}_{k+1}=\Delta p + p^{n+1}_{k}$
           \If {$ \Vert \bu_{k+1}^{n+1}-\bu_k^{n+1}  \Vert < tol $}
                \State \textbf{Update} the Newton solutions $\bu_k^{n+1} \gets \bu_{k+1}^{n+1}$ and $p_k^{n+1} \gets p_{k+1}^{n+1}$
            \Else { $ k \gets k+1$}
           \EndIf
           \State \textbf{Update} mechanical solutions $\bu^{n+1} \gets \bu_{k+1}^{n+1}$ and $p^{n+1} \gets p_{k+1}^{n+1}$
     \EndFor
     \State \textbf{Update time}: $t \gets t+ \Delta t$
     \State \textbf{Output:} Displacement $\bu^{n+1}$, pressure $p^{n+1}$ and electrophysiological variables $(u^{n+1}_s,v^{n+1}_s,u^{n+1}_i, v^{n+1}_i)$ at the current time $t^{n+1}$
\EndWhile
\end{algorithmic}
\end{algorithm}
\newpage

\section{Exploitation of the digital twin model}
\label{sec:simu}
Several numerical experiments are herein presented to characterize the mechanical and electrophysiological response of an idealized tract of the human colon. The present tests aim to investigate and compare colon motility in healthy and post-surgical conditions where a region of the computational domain, corresponding to a lesion in the colon endothelium, has been removed after a representative endoscopic submucosal dissection (ESD) \citep{alnaes2015fenics} and thus replaced with an implant consisting of a patch of 3D printed material. Numerical results are analyzed in terms of topography maps of intraluminal pressure and compared to existing data on colonic high-resolution manometry. Electrophysiological and mechanical constitutive parameters used in the simulations can be found in Tab.s~\ref{table:1} and \ref{table:tm}, respectively.
\begin{table}[h!]
\centering
\caption{Mechanical constitutive parameters.}
\begin{tabular}{c c c c c c c c} 
 \hline
 $\mu \,[\rm kPa ]$ & $k_1^l \,[\rm kPa]$ & $k_2^l \,[-]$ & $k_1^c \,[\rm kPa]$ & $k_2^c \,[-]$ & $k_1^d \,[\rm kPa]$ & $k_2^d \,[-]$ & $\theta \,[-]$\\
 \hline
 $2.5$ & $5.43$ & $1.19$ & $0.78$ & $0.02$ & $3.65$ & $0.31$ & $39.5$\\
 \hline
\end{tabular}
\label{table:tm}
\end{table}
 
\subsection{Idealized colon geometry model}

\begin{figure}[h]
    \centering
    \includegraphics[scale=0.65]{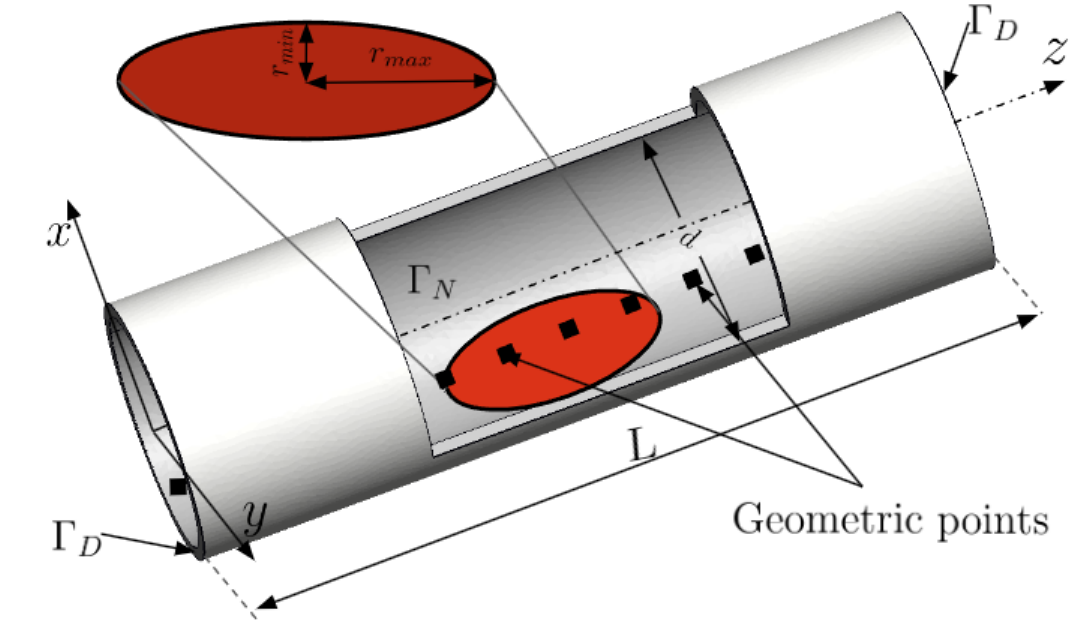}
    \caption{Sketch of the computational domain used in the numerical simulations with length $L=50 \; \rm cm$, diameter $d=5 \; \rm cm$, thickness $0.55 \; \rm cm$. Dirichlet ($\Gamma_D$) and Neumann ($\Gamma_N$) boundary conditions. The red ellipsoidal region with axes $r_{ \rm max}$, $r_{ \rm min}$, and thickness $h=0.3 \, \rm cm$ represents the bio-printed patch.}
    \label{fig:3}
\end{figure}

The computational domain consists of a hollow cylinder with typical dimensions of a human colon geometry. According to colonoscopy and surface data \citep{stauffer2020colonoscopy,helander2014surface}, the diameter of the colon varies between $4.8 \; \rm cm$ and $6\; \rm cm$, while the length of its transverse part is around $50 \; \rm cm$. For the present study, a diameter of $5 \; \rm cm$, a length of $50 \; \rm cm$, and a thickness of $0.5 \; \rm cm$ have been considered. Figure~\ref{fig:3} shows a sketch of the geometry used with associated boundary conditions.

The red elliptic region, with axes $r_{ \rm min}$, $r_{ \rm max}$ and thickness $h=0.3 \; \rm cm$, represents a patch of bio-printed material (e.g., albumin). Two representative geometries for the implanted patch have been considered, fixing the value of the minor axis, $r_{\rm min}=2 \; \rm cm$, and varying $r_{\rm max} \in \{ 2, 3 \} \; \rm cm$. The bio-printed region was modeled using a Neo-Hookean material model, assuming perfect contact between the tissue-patch boundary, thus representing the clinical situation when a healthy bond appears during the healing process. The elastic modulus of the implanted patch, $\mu_p$, has been varied with respect to the elastic modulus of the surrounding healthy tissue, $\mu_t$, as $\mu_p \in \{ \mu_t /2, \mu_t,  2 \mu_t \}$. 

The intraluminal pressure due to mechanical contraction has been evaluated according to Lamé's theory of stresses in thick-walled cylinders \citep{du2013model,choudhury2013stress,lindeburg2019ppi}. Accordingly, a direct comparison with experimental manometry recordings was conducted. The circumferential stresses ($\sigm_c$) in the cylinder wall were calculated using  internal ($p_i$) and external ($p_o$) pressures, and internal ($r_i$) and external ($r_o$) radii of the hollow cylinder, readily:
\begin{equation}
    \sigma_c = \frac{p_i r_i^2-p_o r_o^2}{r_o^2 - r_i^2} - \frac{r_i^2 r_o^2(p_o - p_i)}{r(r_o^2 - r_i^2)} \,,
\end{equation}
where $r$ is the radial coordinate (computed from the Cartesian reference system aligned with the cylinder centerline). For simplicity and with no loss of generality, in the present study, we assumed the external pressure $p_o$ due to surrounding tissues and organs as a null reference value (fulfilling material equilibrium withstanding with colon cylindrical shape). Because the active fibers on the inner surface of the colon are aligned in the circumferential direction, the intraluminal pressure calculated at $r=r_i$ can therefore be expressed as follows:
\begin{equation}
\label{eq:pi}
    p_i = \sigma_c \frac{r_o^2 - r_i^2}{r_o^2 + r_i^2} \,,
\end{equation}
such that, after solving electromechanical equilibrium, $\sigma_c$ is known and the intraluminal pressure can be estimated.

The topographic maps of intraluminal pressure simulating high-resolution manometry were created according to the clinical protocol described in \citep{conklin2009color,li2019high}: the intraluminal pressure has been evaluated at 36 geometric points along the computational domain in the numerical simulations; after computing $p_i$ values, an in-house code was written in \texttt{Matlab2022} to display the results of the simulations.

\subsection{The role of implant material on intraluminal pressure}

\begin{figure}[ht!] 
\centering
\includegraphics[width=\textwidth]{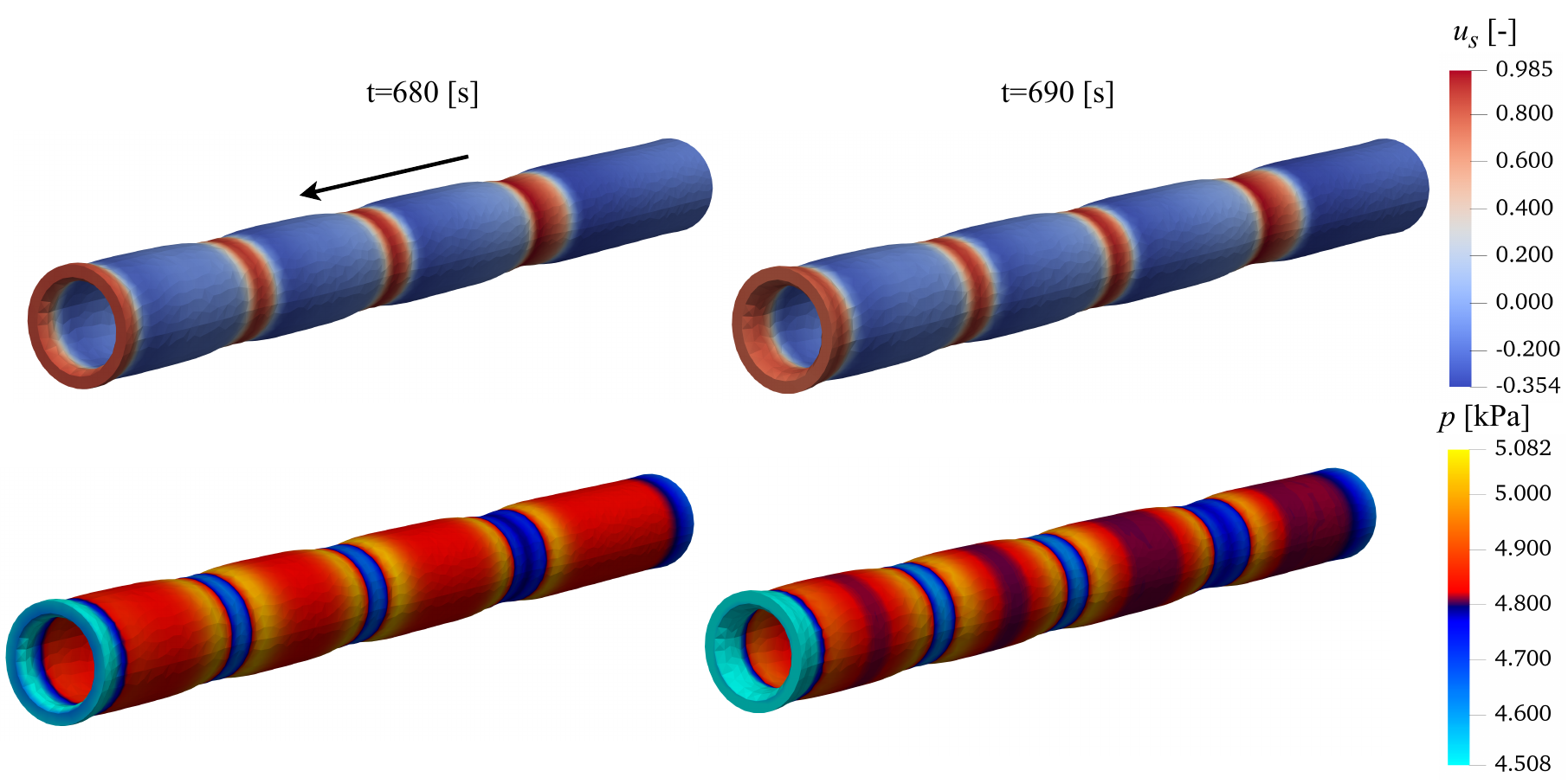}
\\
\includegraphics[width=0.98\textwidth]{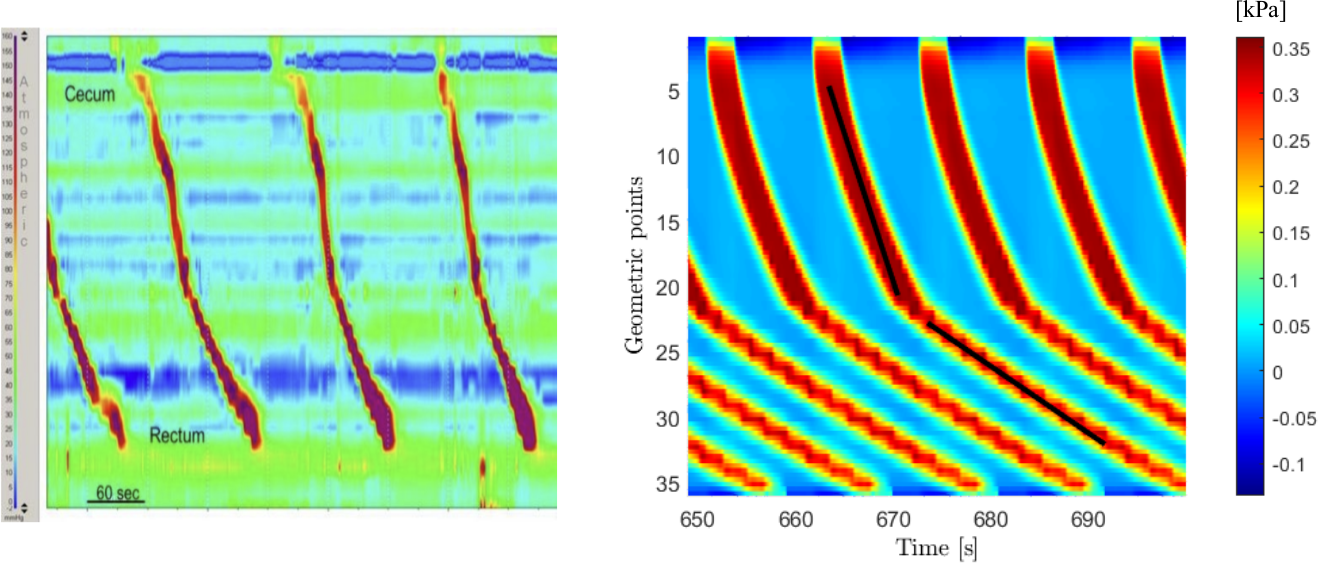}
\caption{
(Top) Temporal evolution of the SMC transmembrane potential $u_s$ and hydrostatic pressure $p$ in the healthy condition ($\mu_p=\mu_t$). The arrow represents the direction of propagation.
(Bottom) Topography map of the intraluminal pressure $p_i$ corresponding to HRM map in a healthy colon tract: (a) clinical results taken from \citep{arbizu2017prospective}, (b) numerical model with $\mu_p=\mu_t$. Black lines represent the slope, i.e., conduction velocity, in the space-time diagram.}
\label{fig:4h}
\end{figure}

\paragraph{Healthy case}
The first numerical test is aimed to simulate the electromechanical behavior of a healthy colonic tract (corresponding to the condition $\mu_p=\mu_t$). Figure \ref{fig:4h} shows the numerical results obtained by finite element simulations for two representative snapshots (a zoomed clip of the region of interest is shown in Fig.~\ref{fig:4}). In particular, the spatial distribution of SMC transmembrane potential $u_s$ (first row) and hydrostatic pressure $p$ (second row) are shown on the deformed/contracted state. According to the active strain constitutive modeling assumptions \eqref{eq:2}, the deformation is in phase with the peak of slow wave activity. After a transient period of system stabilization (see \ref{sec:A} for details), multiple slow waves coexist on the domain and propagate in the axial direction, correctly reproducing healthy colon peristalsis.

\begin{figure}
    \centering
    \includegraphics[width=0.65\textwidth]{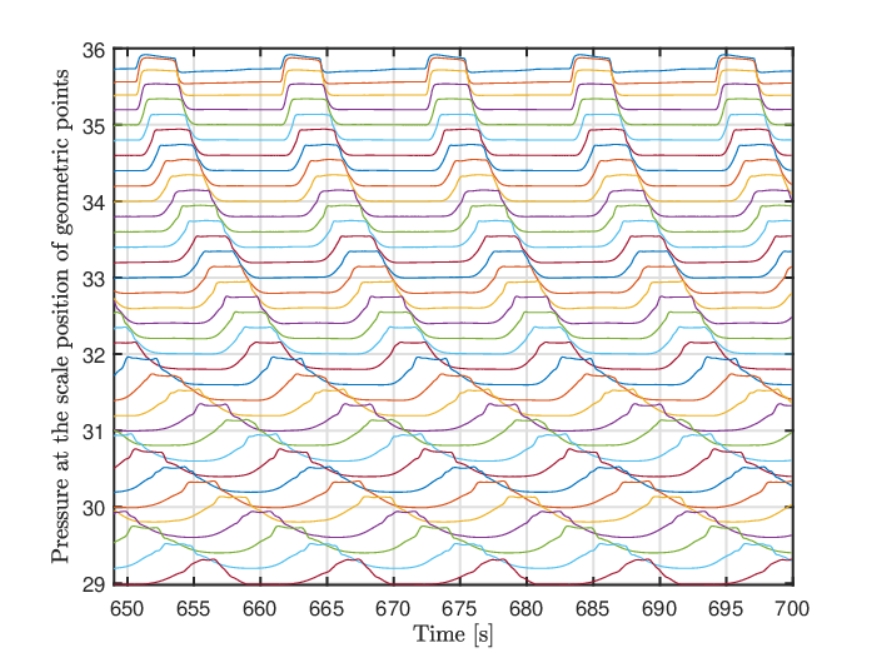}
    \caption{Pressure topography map corresponding to the numerical manometry in Fig. \ref{fig:4h}.}
    \label{fig:healthyTopo}
\end{figure}
Besides, manometry maps (Fig.~\ref{fig:4h}) and pressure topography maps (Fig.~\ref{fig:healthyTopo}) provide a faithful representation of the intraluminal pressure generated by muscle contraction. Manometry map analysis reveals qualitative agreement with clinical data from the literature \cite{arbizu2017prospective}, as illustrated in Fig.~\ref{fig:4h} for the healthy case. Specifically, the space-time diagram highlights the intensity and speed of propagation (slope computed as space/time) of $p_i$ field \eqref{eq:pi}. As observed in clinical data, a stronger intraluminal pressure is measured according to the excitation wave speed, thus changing along the GI tract in favor of the mixing function. Moreover, it is worth mentioning that in the healthy case, the motility pattern follows the propagating waves smoothly without impediments or gaps.



\newpage
\paragraph{The role of patch geometry and stiffness}
In this section, we provide a preliminary parametric analysis comparing different properties of the bio-printed patch, namely geometry and stiffness. 
\begin{figure}[ht!] 
\centering
{\includegraphics[width=0.92\textwidth]{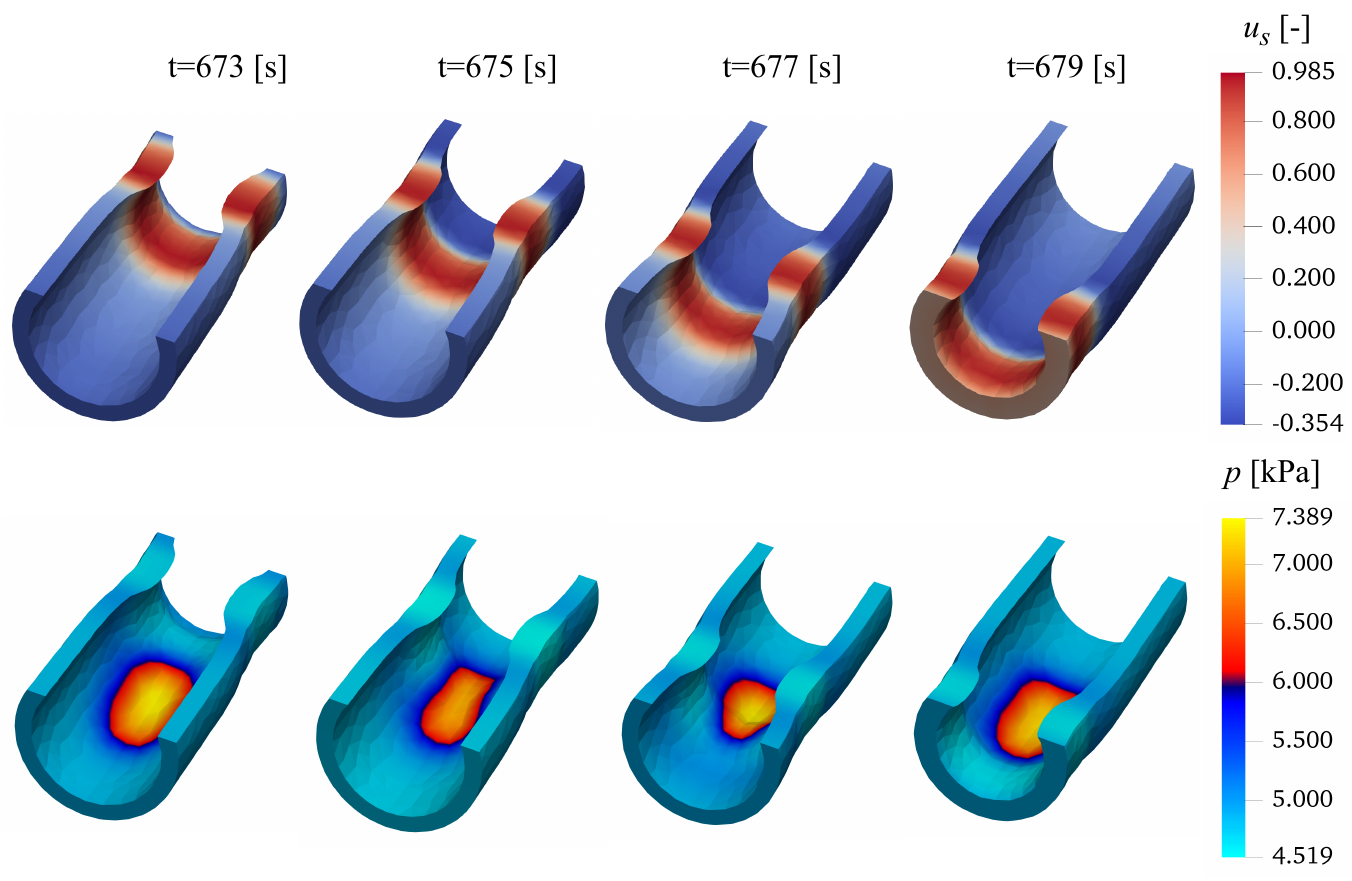}}
\hspace{0.005\textwidth}
{\includegraphics[width=0.92\textwidth]{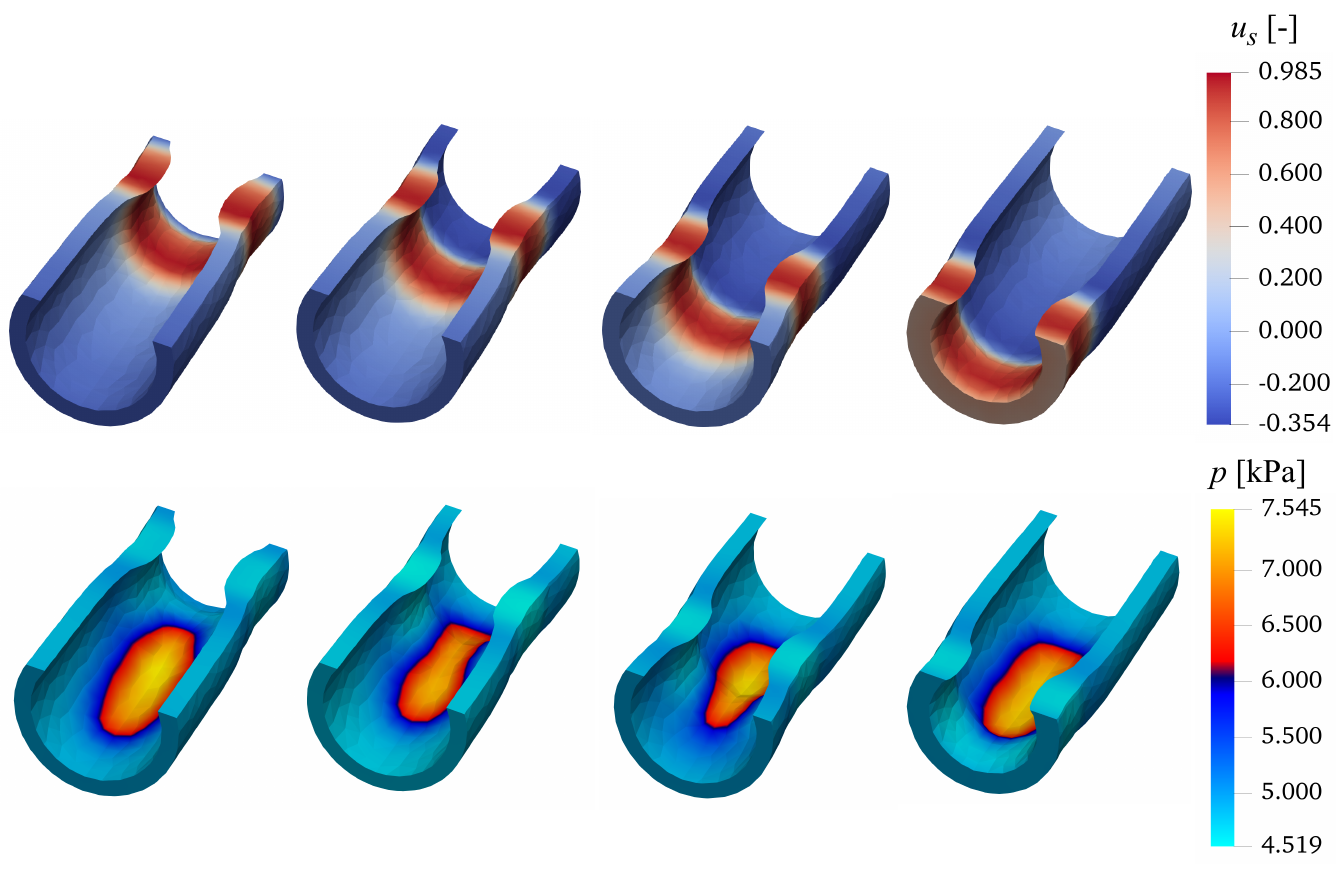}}
\caption{Temporal evolution of SMC transmembrane voltage $u_s$ and hydrostatic pressure $p$ in proximity of the patch region with parameters $r_{\rm max}= 2,r_{\rm min}=2,\mu_p=2 \mu_t$ (top), and $r_{\rm max}= 3,r_{\rm min}=2,\mu_p=2 \mu_t$ (bottom).}
\label{fig:6}
\end{figure}
Figure \ref{fig:6} shows a graphical representation of the numerical results obtained varying the size of the elliptical region $r_{\rm max} \in \{2, 3\} \; \rm cm$ and considering a stiffer patch material other than the surrounding tissue, namely $\mu_p= 2 \mu_t$.

It can be noticed that although the slow wave propagation is not considerably altered by the presence of the patch (representative of the case when the implant is fully cellularized), such that only geometrical couplings are implied, the intraluminal pressure distribution critically changes due to the presence of a stiff patch. In particular, the overall patch area shows much higher $p_i$ levels (from about 5 kPa in the healthy case to more than 7 kPa in the stiff patch case), concurring to increase the stress state of the surrounding tissue and thus enhancing the development of pathological scars \citep{gizzi2012,pierfranco2018}.


\begin{figure}[ht!] 
\centering
\includegraphics[width=0.93\textwidth]{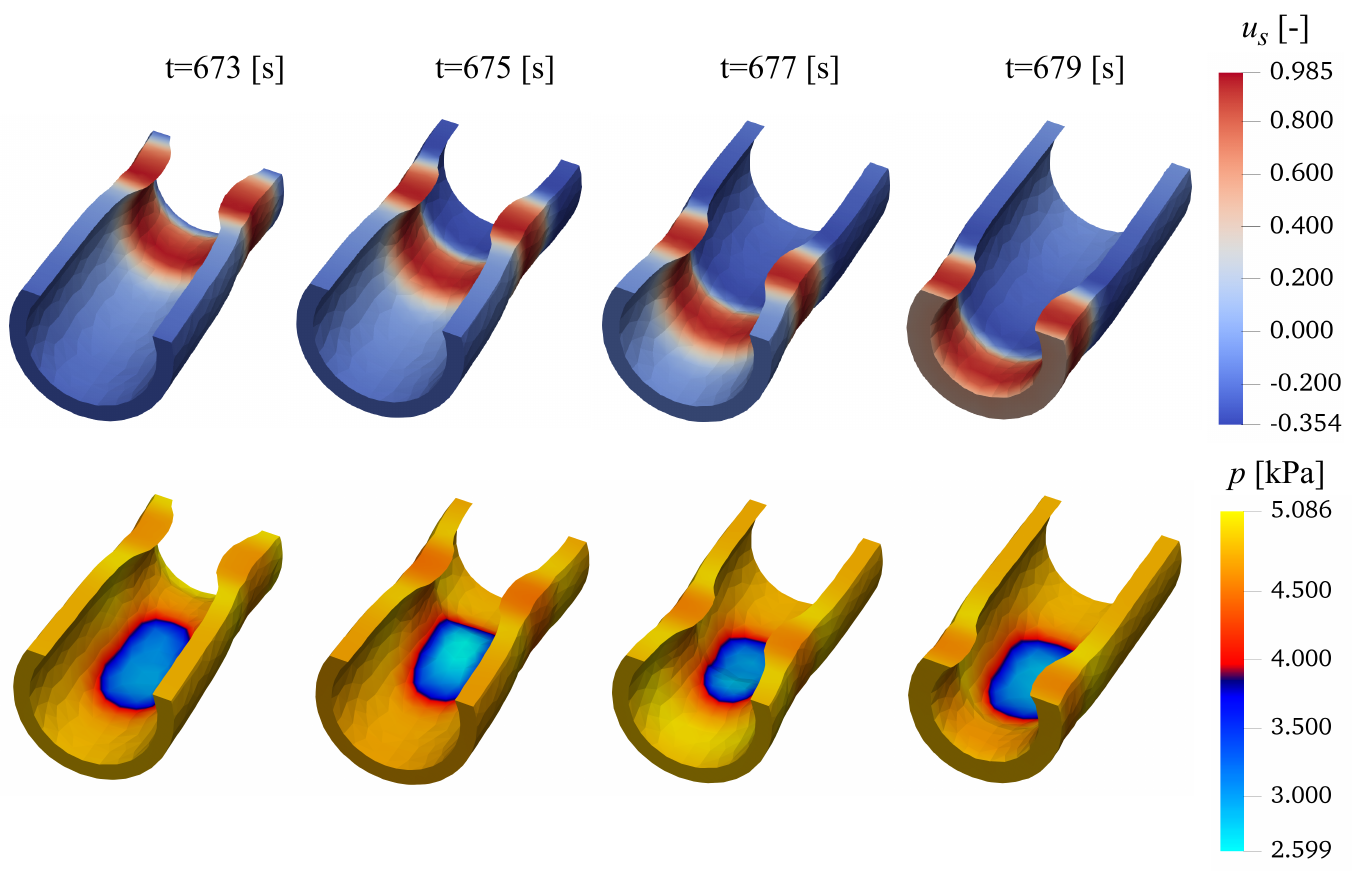}
\hspace{0.005\textwidth}
\includegraphics[width=0.93\textwidth]{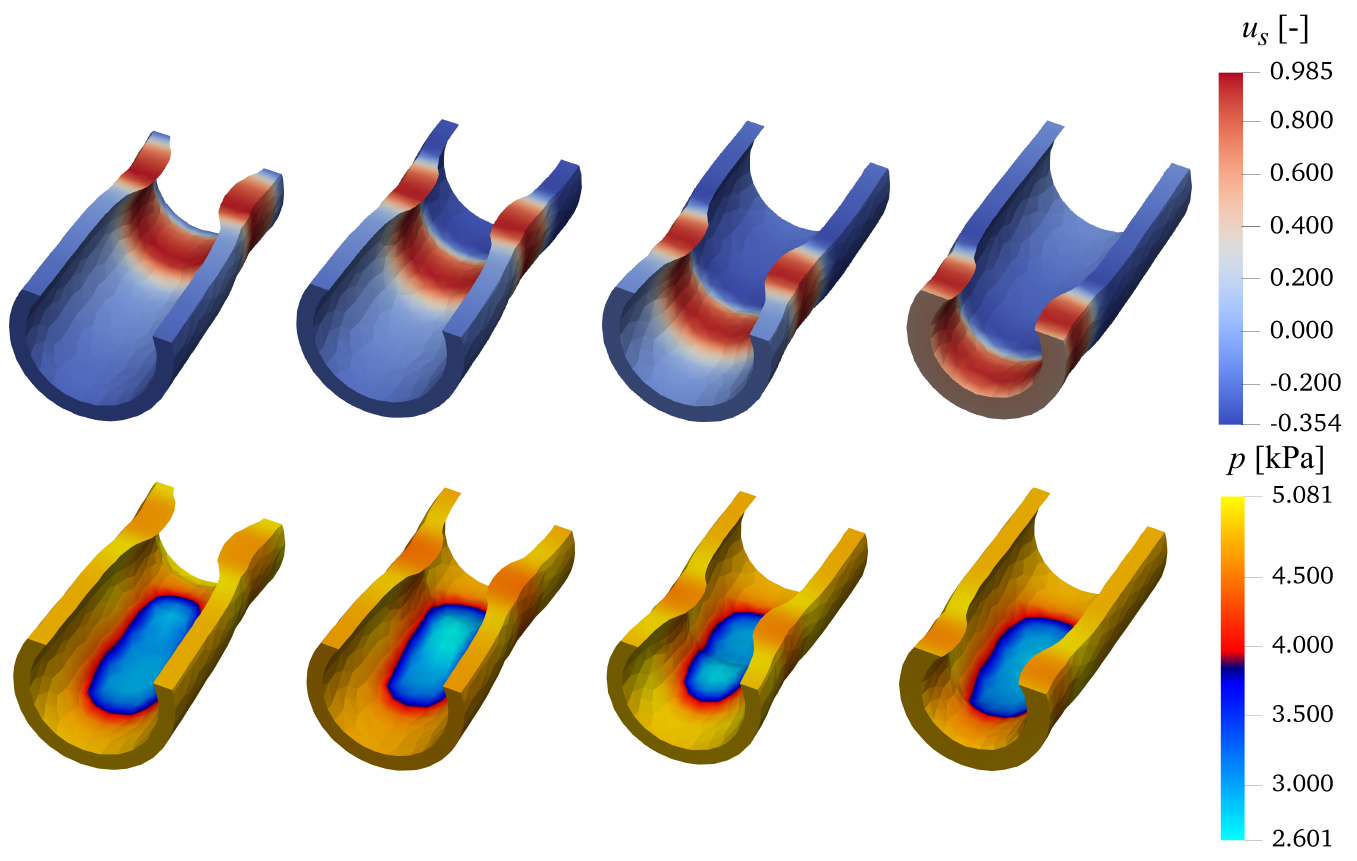}
\caption{Temporal evolution of SMC transmembrane voltage $u_s$ and hydrostatic pressure $p$ in proximity of the patch region with parameters $r_{\rm max}= 2,r_{\rm min}=2,\mu_p=0.5 \mu_t$ (top), and $r_{\rm max}= 3,r_{\rm min}=2,\mu_p=0.5 \mu_t$ (bottom).}
\label{fig:8}
\end{figure}

Figure \ref{fig:8} shows analyses conducted varying patch size $r_{\rm max} \in \{2, 3\} \; \rm cm$ but considering a softer material other than the host surrounding tissue, namely $\mu_p= \mu_t / 2$. As expected, a similar behavior is obtained for the SMC voltage field, i.e., the voltage field only experiences geometric coupling, but, in this case, the intraluminal pressure lowers (from about 5 kPa in the healthy case to less than 3 kPa in the soft patch case) on the overall patch region. Such a general stress mismatch could affect the stability of the implant (it may lose its position) with concurrent loss of contractility efficiency.

\begin{figure}[ht!] 
\centering
\includegraphics[width=1\textwidth]{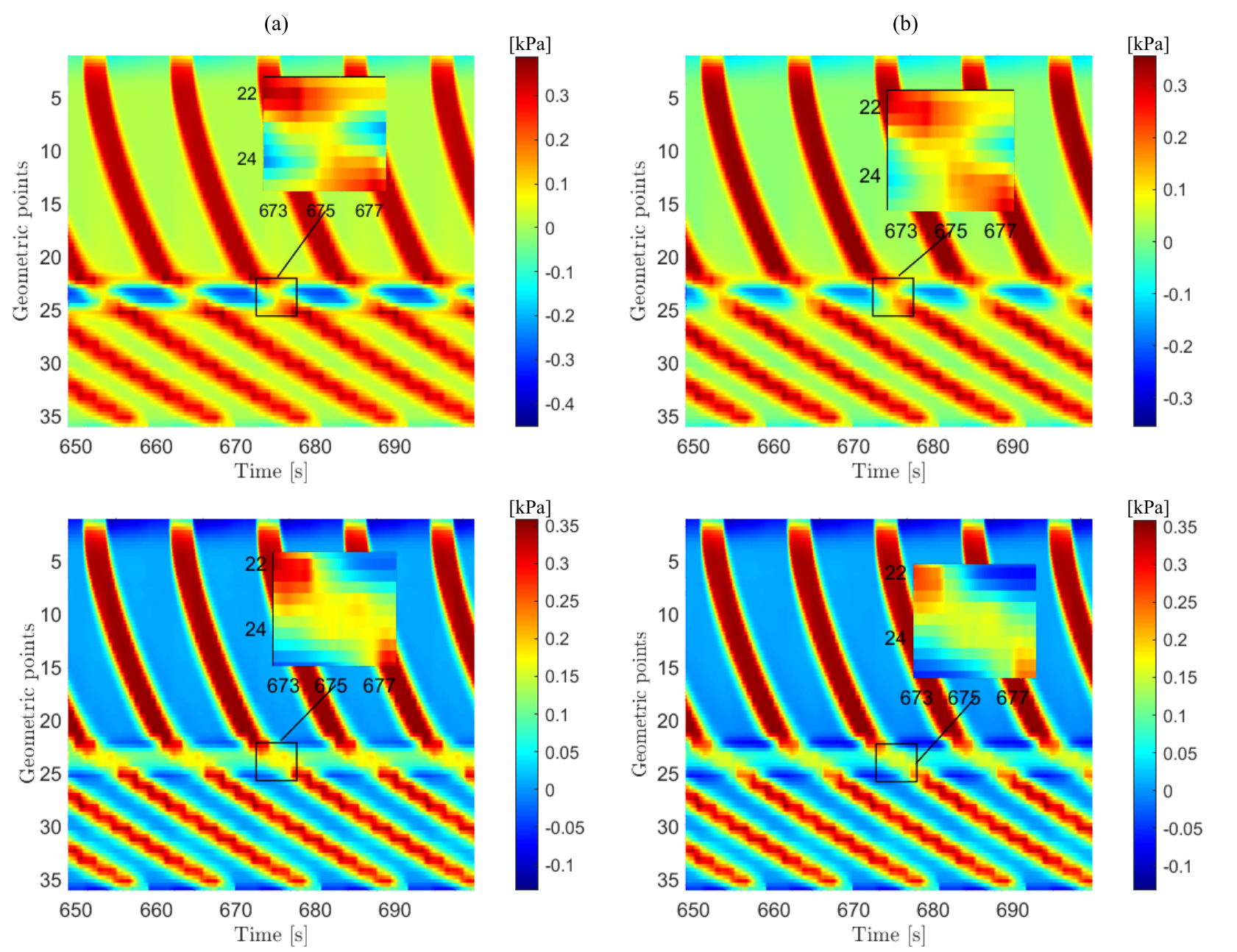}
\caption{HRM maps for two implant configurations after the LTS with parameters $\mu_p=2 \mu_t$ (top) and $\mu_p=0.5 \mu_t$ (bottom). Columnwise discriminates between $r_{\rm max}= 2,r_{\rm min}=2$ (a) and $r_{\rm max}= 2,r_{\rm min}=3$ (b).}
\label{fig:10}
\end{figure}

\begin{figure}[ht!] 
\centering
\includegraphics[width=1\textwidth]{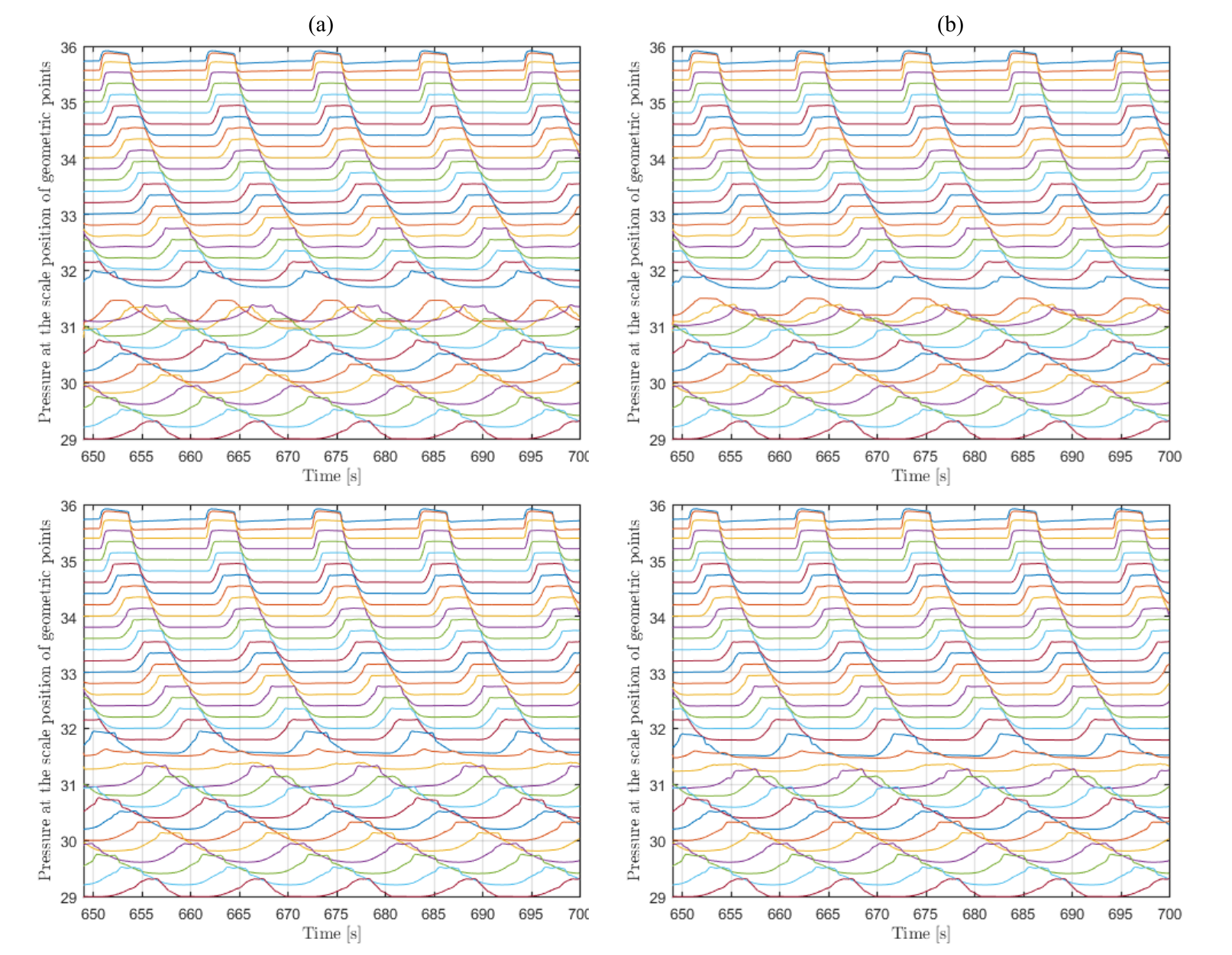}
\caption{Topography pressure maps for the HRM in Fig. \ref{fig:10} two implant configurations after the LTS with parameters $\mu_p=2 \mu_t$ (top) and $\mu_p=0.5 \mu_t$ (bottom). Columnwise discriminates between $r_{\rm max}= 2,r_{\rm min}=2$ (a) and $r_{\rm max}= 2,r_{\rm min}=3$ (b).}
\label{fig:Topomappatch}
\end{figure}

To further emphasize the critical role of material stiffness in the overall colon motility, we discuss the resulting topography maps for the selected pathological cases, as shown in Fig.~\ref{fig:10} and Fig.~\ref {fig:Topomappatch}. A direct comparison with the healthy case provided in Fig.~\ref{fig:4h} highlights that the region surrounding the patch material reduces its contractility for a stiffer or softer patch. Moreover, a critical gap can be observed in the intraluminal pressure profiles that break, thus reducing the lateral wall displacement. Such a contractility impairment is more evident in the case of the stiff patch, concurring to a wrong scar remodeling in the long-term tissue adaptation. Contrarily, selected differences in patch size do not alter the resulting motility behavior, which seems to be ruled by the patch material properties other than its geometrical features.


\newpage
\paragraph{The role of patch electrical conductivity}

In this numerical test, the electrophysiological properties of the implant are varied from those of the surrounding host tissue; namely, the SMC and ICC electrical conductivity in the patch, $D^p_s, D^p_i$, are ten times lower than those in the tissue, $D^t_s, D^t_i$. The size of the elliptical region is fixed at $r_{\rm min}=2 \, \rm cm$ and $r_{\rm max}=3 \, \rm cm$ and material stiffness is considered $\mu_p=2\mu_t$.

\begin{figure}[ht!] 
\centering
\includegraphics[width=0.90\textwidth]{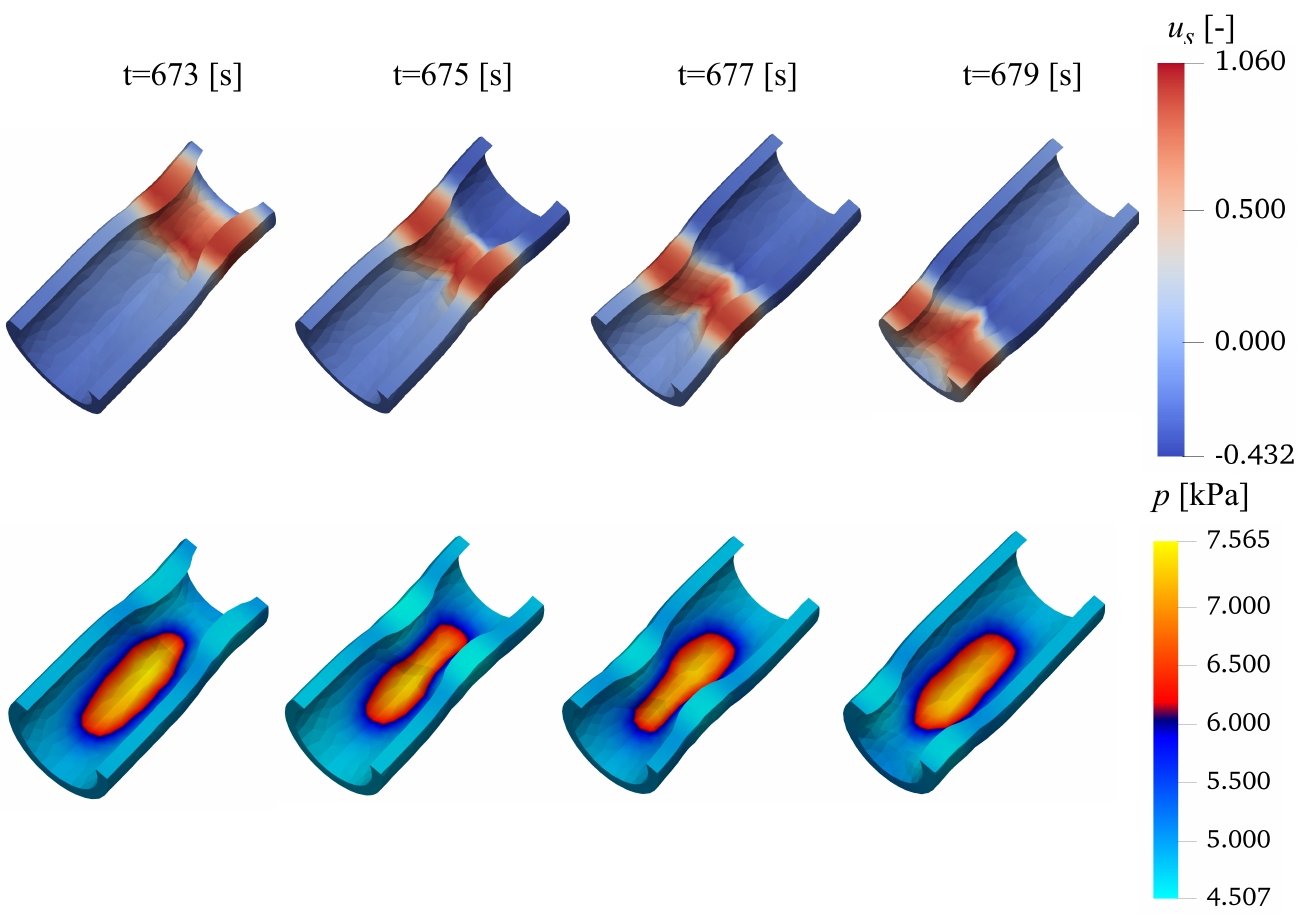}
\\
\includegraphics[width=\textwidth]{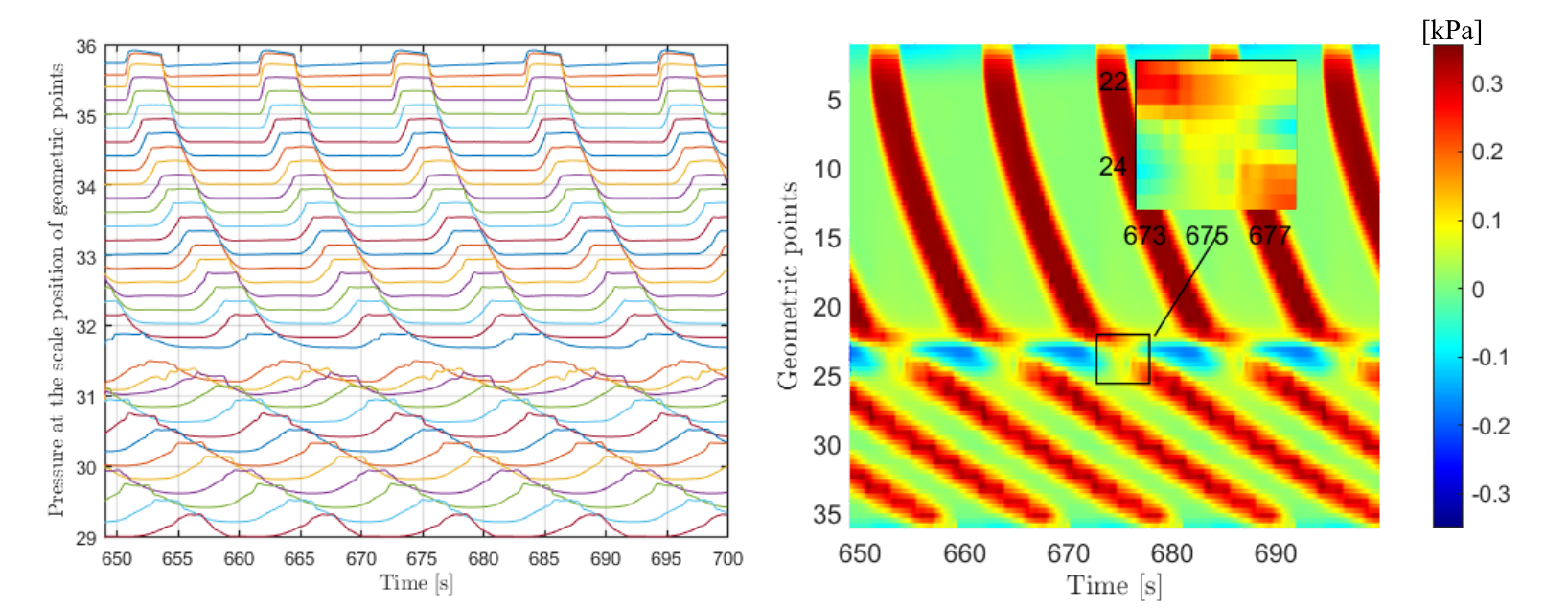}
\caption{Temporal evolution of SMC transmembrane voltage $u_s$ and hydrostatic pressure $p$ around the patch region with parameters $r_{\rm max}= 3$ and $r_{\rm min}=2$ with stiffness $\mu_p=2 \mu_t$ and the diffusion coefficients $D_s^p=0.1D_s$ and $D_i^p=0.1D_i$.
HRM map  with $\mu_p=2 \mu_t$ with in-homogeneous diffusivity $D_s^p=0.1D_s$ and $D_i^p=0.1D_i$.}
\label{fig:11}
\end{figure}

Figure \ref{fig:11} shows the evolution of the hydrostatic pressure $p$ in the surroundings of the implant, where no significant variations can be observed compared to the previous analysis in Fig.~\ref{fig:6}. However, because of the altered electrophysiological properties, the slow wave spatiotemporal distribution changes, and, in particular, the conduction velocity of the excitation wave lowers by enlarging the action potential wavelength when passing across the implant. Such a perturbation affects the overall displacement field inducted on the colon surrounding the patch, concurring with an altered pattern obtained for the intraluminal pressure map. Accordingly, such a preliminary analysis, in conjunction with the topographic pressure profiles in Fig.~\ref{fig:11}, confirms the sensitivity of colon motility to material stiffness and excitability. The gap in the $p_i$ map is more pronounced than the case in Fig.~\ref{fig:10} because the contraction of the tissue surrounding the patch is delayed. To better understand the effect of diffusivity, we repeated the simulation by reducing again the diffusivity of the patch by ten, and the results are presented in \ref{sec:E}. The results confirm no significant difference with those provided in Fig.~\ref{fig:11}, which means that diffusivity alone has a minor effect on the overall material behavior.

\paragraph{The role of patch contractility}
In this numerical test, the implant electrophysiological properties and contractility vary from the surrounding host tissue. Namely, the SMC and ICC electrical conductivity in the patch, $D^p_s, D^p_i$, are a hundred times lower than those in the tissue, $D^t_s, D^t_i$, and the amplitude of contractility in Eq.~\eqref{eq:2}, $\alpha_l^p, \alpha_c^p$, is reduced as by  $50\%$. The size of the elliptical region is fixed at $r_{\rm min}=2 \, \rm cm$ and $r_{\rm max}=3 \, \rm cm$ and material stiffness is considered $\mu_p=2\mu_t$.

\begin{figure}[ht!] 
\centering
\includegraphics[width=\textwidth]{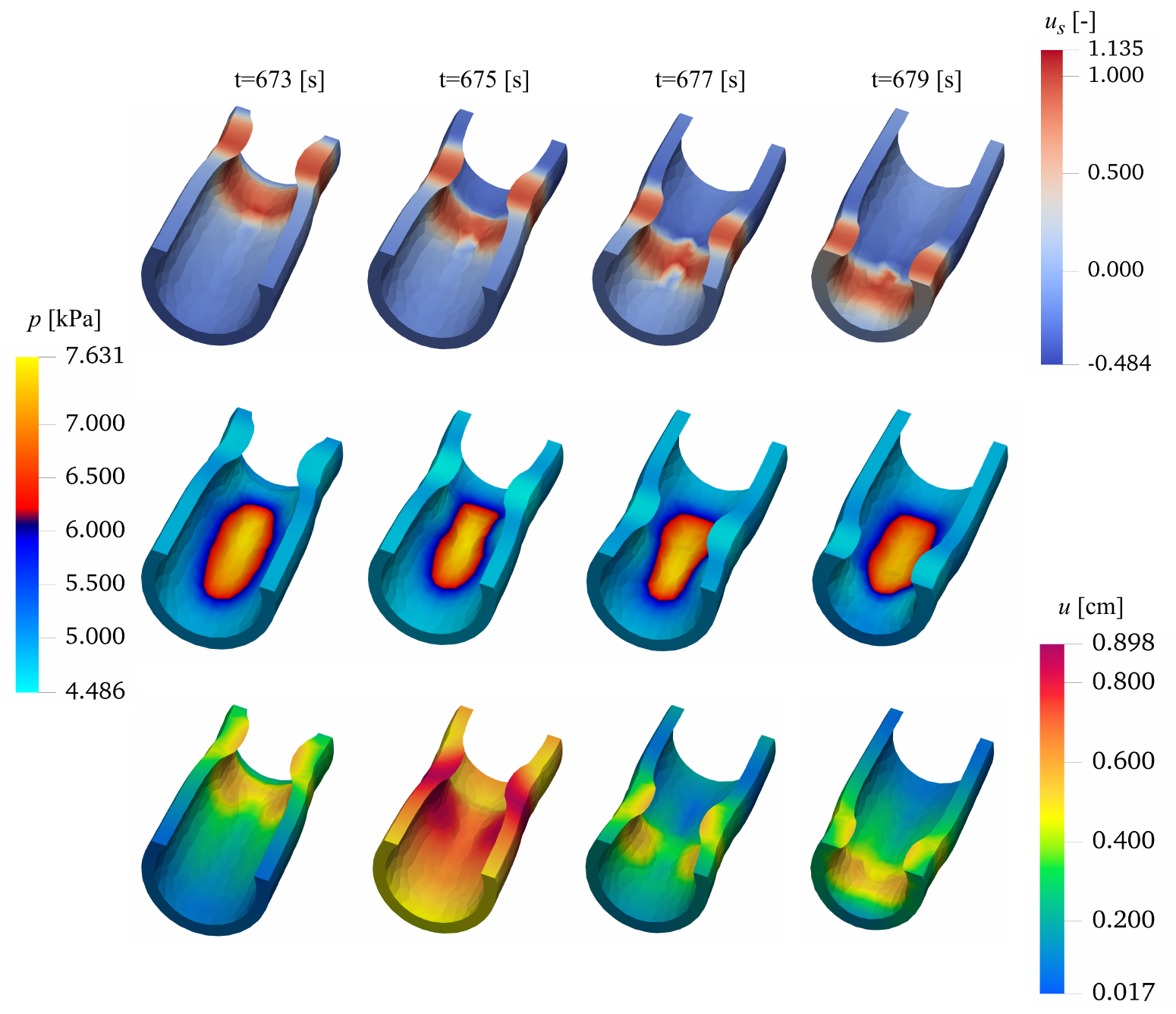}
\\
\includegraphics[width=\textwidth]{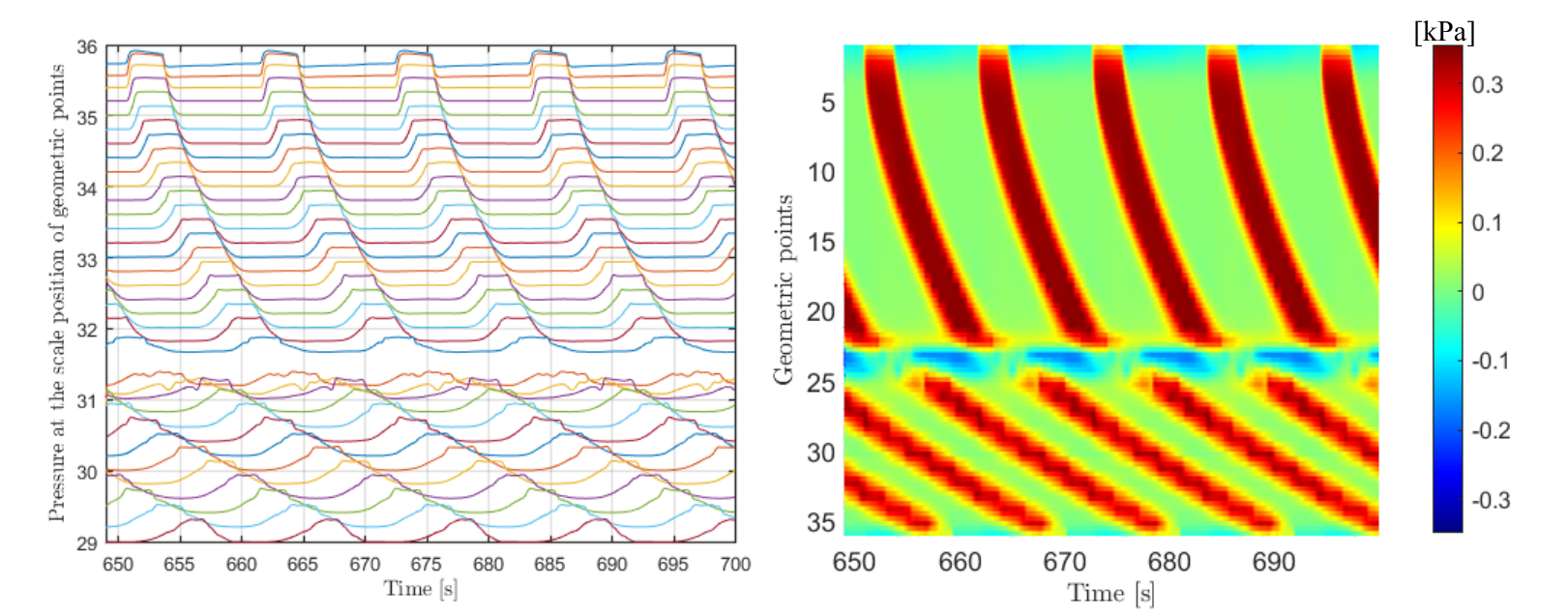}
\caption{Temporal evolution of SMC transmembrane voltage $u_s$, of hydrostatic pressure $p$ and the displacement $u$ around the patch region with parameters $r_{\rm max}= 3$ and $r_{\rm min}=2$ with stiffness $\mu_p=2 \mu_t$ and the diffusion coefficients $D_s^p=0.01D_s$ and $D_i^p=0.01D_i$.
HRM map  with $\mu_p=2 \mu_t$ with in-homogeneous diffusivity $D_s^p=0.01D_s$ and $D_i^p=0.01D_i$ and the contractility $\alpha_l^p = 50\% \alpha_l$ and $\alpha_c^p = 50\% \alpha_c$.}
\label{fig:NonDiffContra}
\end{figure}

Figure \ref{fig:NonDiffContra} shows the evolution of the hydrostatic pressure $p$ in the surroundings of the implant. It is worth noting that due to the reduction of contractility, significant variations can be observed in hydrostatic pressure compared to the case in Fig.~\ref{fig:6}. As electrophysiological properties are altered, the slow wave spatiotemporal distribution changes, and, in particular, the conduction velocity of the excitation wave lowers by enlarging the action potential wavelength when passing across the implant. The two contributions notably affect the overall displacement field $u$ where a low displacement of 0.4 cm is observed around the patch (see third row in Fig.~\ref{fig:11}), enforcing an altered motility pattern demonstrated by a clear gap in the intraluminal pressure profile in Fig.~\ref{fig:NonDiffContra}--a $p_i$ gap more pronounced compared to the case in Fig.~\ref{fig:11}. Accordingly, such analysis confirms the sensitivity of colon motility to material stiffness, excitability, and contractility. 
\clearpage
\newpage
\paragraph{The role of a non-cellularized patch}
In this numerical test, we consider a patch not yet fully cellularized (representative of an early healing stage) with altered electrophysiological properties. Namely, the SMC and ICC electrical conductivity $D^p_s, D^p_i$ are $10^2$ times lower than those in the tissue, $D^t_s, D^t_i$, and the reaction terms Eq.~\eqref{eq:19} are $10^3$ times lower (i.e., the reaction terms are multiplied by a constant $10^{-3}$ factor). The patch size and stiffness are maintained as in the previous case.

\begin{figure}[ht!] 
\centering
\includegraphics[width=0.95\textwidth]{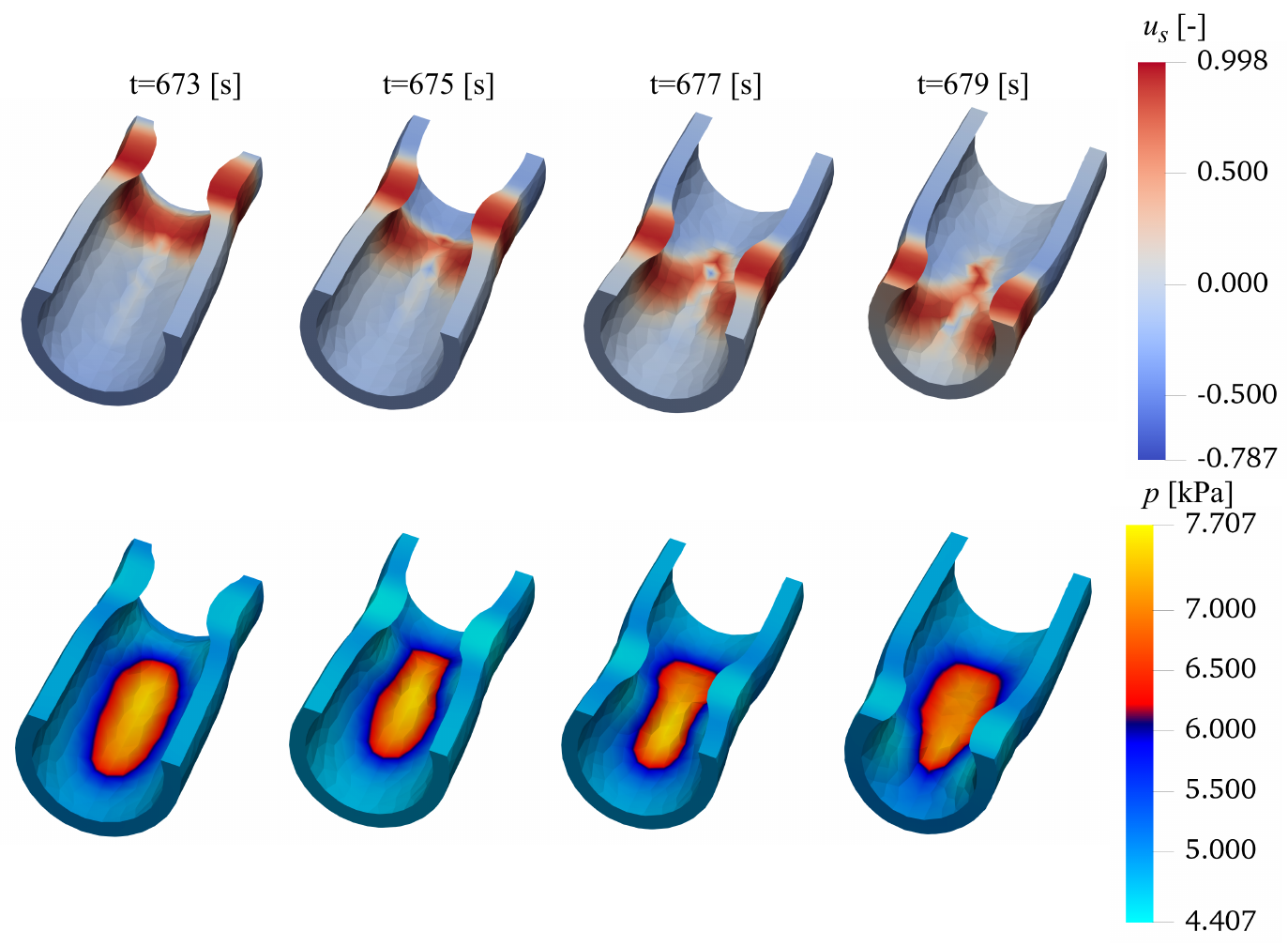}
\\
\includegraphics[width=\textwidth]{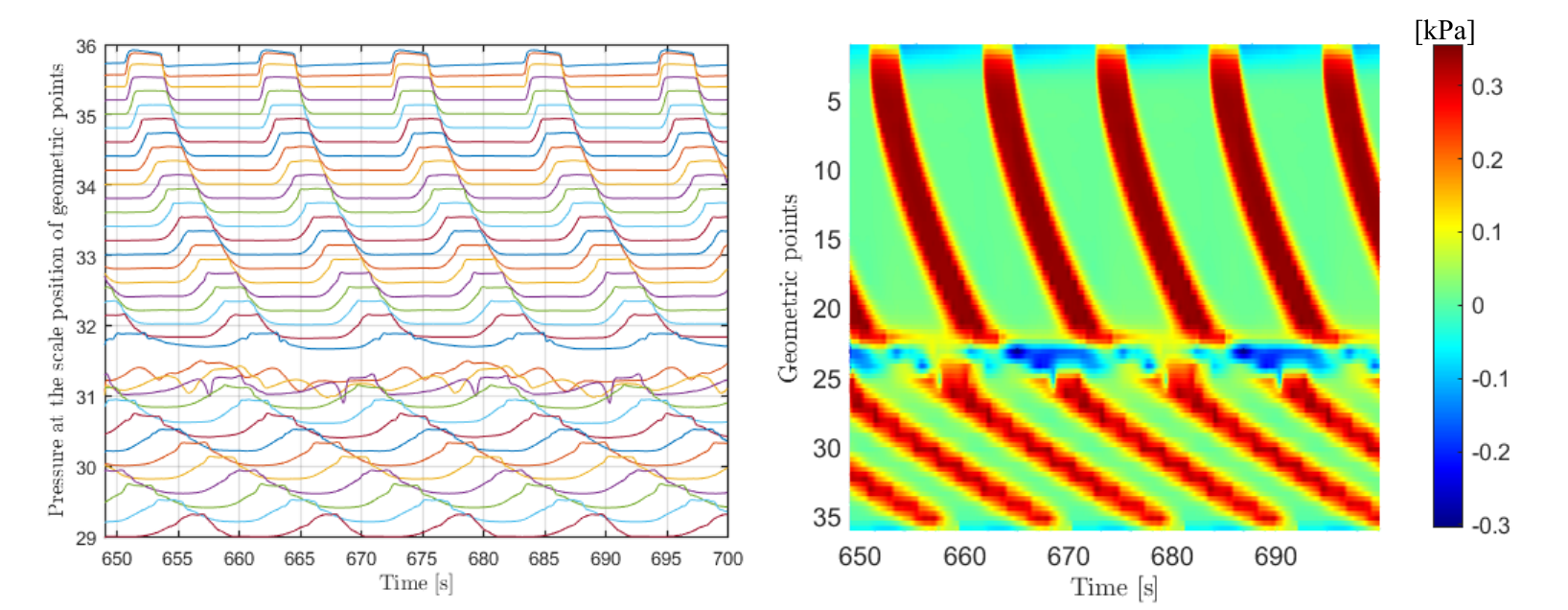}
\caption{Temporal evolution of SMC transmembrane voltage $u_s$ and hydrostatic pressure $p$ around the patch region with parameters $r_{\rm max}= 3$ and $r_{\rm min}=2$ with stiffness $\mu_p=2 \mu_t$ and the diffusion coefficients $D_s^p=0.01D_s$ and $D_i^p=0.01D_i$.
HRM map with $\mu_p=2 \mu_t$ with in-homogeneous diffusivity $D_s^p=0.01D_s$ and $D_i^p=0.01D_i$ with the reaction terms (Eq.~\eqref{eq:19}) $10^3$ times lower inside the patch.}
\label{fig:NonDiffReact}
\end{figure}

Figure \ref{fig:NonDiffReact} shows the evolution of the hydrostatic pressure $p$ in the surroundings of the implant. Though no significant variations are observed compared to the case in Fig.~\ref{fig:6}, altered electrophysiological properties induce a critical change in slow wave spatiotemporal dynamics leading to wavebreaks around the patch (no closed rings). In particular, the excitation wave is slowed, and its amplitude is reduced, thus flattening and prolonging the duration of SMC action potential. The two reaction-diffusion contributions affect the colon wall overall motility, leading to an altered pattern of intraluminal pressure maps: a clear gap in the recorded traces is representative of no contraction. Such analysis is aligned with the topographic pressure profiles presented in Fig.~\ref{fig:NonDiffReact}, further confirming the sensitivity of colon motility to bio-printed material cellular viability.

\clearpage
\newpage
\paragraph{The role of patch contractility and altered electrophysiology}
In the last numerical test, we combine the multiple cases discussed before by considering a not fully cellularized patch (reducing the reaction terms by $10^{-3}$ and the diffusion coefficients by $10^{-2}$) but also considering a reduced material contractility via $\alpha_l^p, \alpha_c^p$ lowered by $50\%$. Size and stiffness of the patch are maintained as in the previous case.

\begin{figure}[ht!] 
\centering
\includegraphics[width=\textwidth]{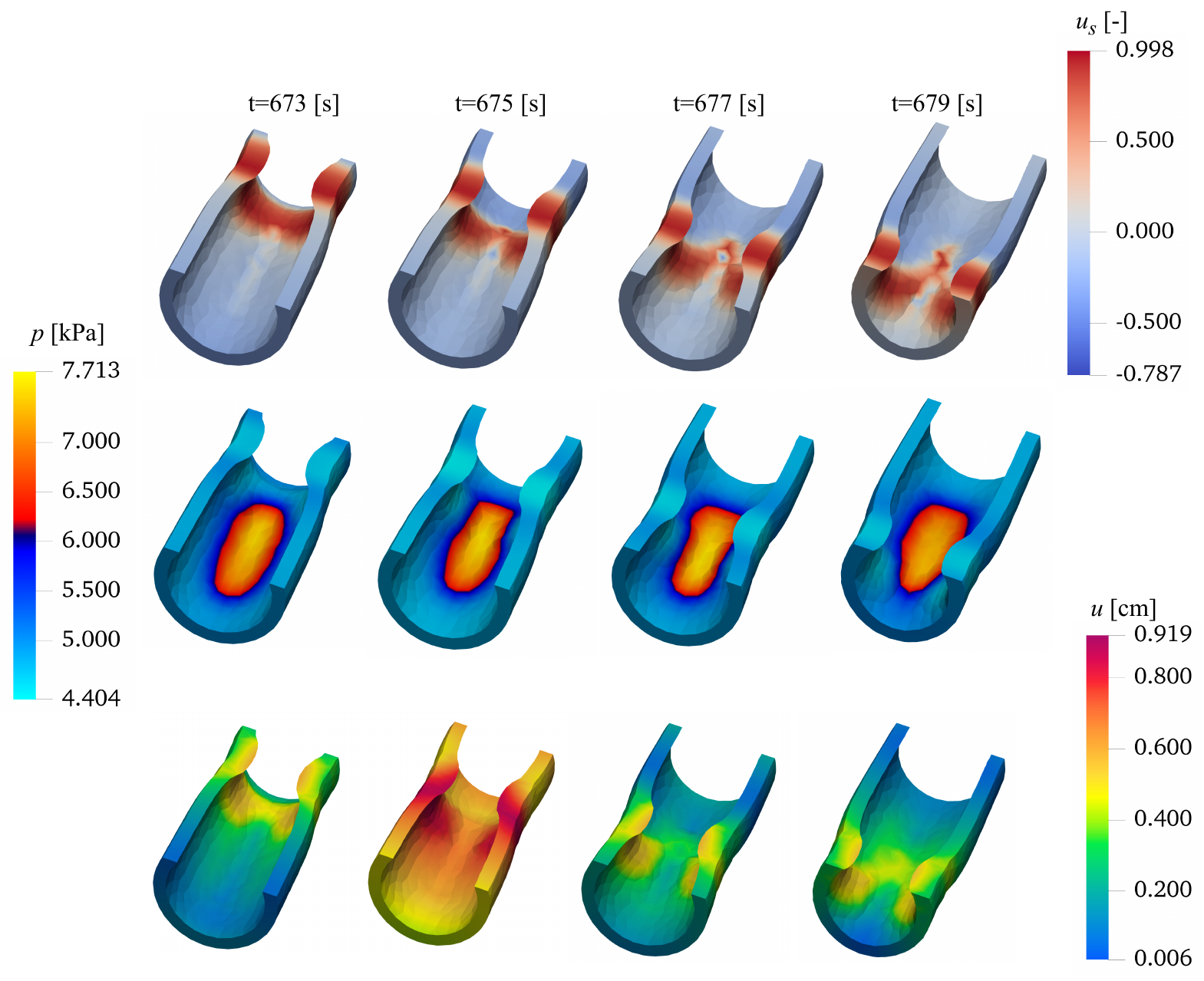}
\\
\includegraphics[width=\textwidth]{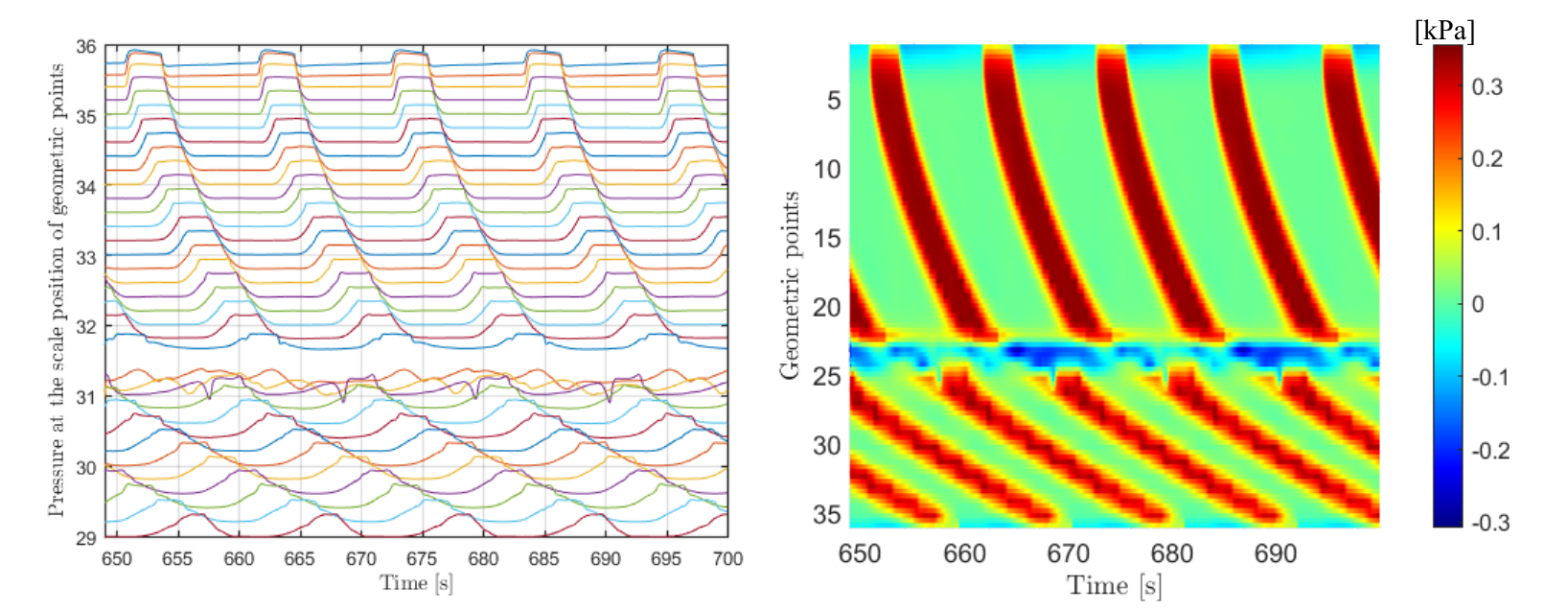}
\caption{Temporal evolution of SMC transmembrane voltage $u_s$, the hydrostatic pressure $p$ and the displacement $u$ around the patch region with parameters $r_{\rm max}= 3$ and $r_{\rm min}=2$ with stiffness $\mu_p=2 \mu_t$ and the diffusion coefficients $D_s^p=0.01D_s$ and $D_i^p=0.01D_i$.
HRM map and the pressure map with $\mu_p=2 \mu_t$ with in-homogeneous diffusivity $D_s^p=0.01D_s$ and $D_i^p=0.01D_i$ and the contractility $\alpha_l^p = 50\% \alpha_l$ and $\alpha_c^p = 50\% \alpha_c$ and the reaction terms (Eq.~\eqref{eq:19}) $10^3$ times lower inside the patch.}
\label{fig:NonDiffReactContracdis}
\end{figure}

Figure \ref{fig:NonDiffReactContracdis} shows the results of the combined ill cases. As expected, hydrostatic pressure and slow waves are affected similarly, as well as the displacement field $u$, which is practically two times lower than the case in Fig.~\ref{fig:NonDiffContra}. However, regarding the HRM and topography maps, no contraction is associated with a reduced displacement to $0.006$ cm complemented with a larger gap in the $p_i$ map comprising a wider region around the patch. Such a comprehensive analysis confirms the role and sensitivity of colon motility to bio-printed material properties in various features. 
\clearpage
\newpage

\section{Conclusion}
\label{sec:con}

Several factors still hinder the advancement of GI electromechanics. These include the intricate and historically poorly understood GI electrophysiology, the complex mechanics of digesta (food undergoing digestion) tightly coupled to its electrical features, and the limited knowledge of neural and hormonal mechanisms regulating GI motility. Over the past decade, significant strides have been made in all these areas, and the exponential growth in computational power now enables tackling even intricate multiphysics problems.

This paper presented a comprehensive multi-field computational framework to model colonic motility, incorporating active strain electromechanics, tissue anisotropy, and cellular electrophysiology. The obtained digital twin has been exploited to characterize colon contractility in the presence of bio-printed deposited materials, e.g., by LTS procedures, assumed to bond perfectly to the surrounding healthy tissue. 

The fully coupled system was numerically solved via a custom finite element staggered scheme to exchange information among the electrophysiological and mechanical solvers efficiently. An extended calibration activity was preliminarily conducted to provide a robust and reliable digital twin model, thus performing several benchmark tests: 
(i) the study of transmembrane action potential conduction velocity;
(ii) the entrainment frequency and stability of ICCs and SMCs temporal dynamics;
(iii) the fine-tuning of strain energy material parameters upon multiaxial experimental data. 

As an additional modeling step, the numerical values of intraluminal pressures generated by colon contractions have been compared to high-resolution manometry patterns, showing qualitative and quantitative agreement with clinical data and providing a novel in silico characterization of colon motility in health and disease. Accordingly, a series of parametric numerical analyses revealed that excessive or reduced patch stiffness could affect the capability of the colon to generate effective muscle contractions, resulting in impaired motility. Furthermore, altered electrophysiological properties, also connected with muscular contractility ruled by the active deformation gradient, confirmed the critical role of a cellularized patch. In particular, the proposed digital twin is able to identify the possible outcomes of altered motility due to LTS at different stages of the patch healing process.

\subsection{Limitations and perspectives}
The present model can reproduce contraction patterns of a colon tract by embedding one single type of contraction wave, either high (HAPCs) or low (LAPCs) amplitude propagation contractions \citep{dinning2015use,li2019high}. Such a limitation is linked to the physiological onset of excitation waves, often initiated by the enteric nervous system (ENS), not modeled in the present work \citep{fung2020functional,barth2017electrical,barth2018computational}. In a forthcoming study, we are considering extending the present formulation to account for ENS in a reliable and robust numerical implementation.

In the present work, we considered a perfectly bounded patch representative of a correct healing process. Though not exploited, the digital twin has already been formulated to account for generalized contact mechanics problems. In particular, irregular or localized faults can be considered in future studies.

In view of a preliminary parametric analysis and the lack of dedicated data, we assumed uniform external pressure representative of an elastic and isotropic surrounding tissue. A possible extension of the study should also consider a cohesive zone model and material remodeling to reproduce the multiple interfaces and loading acting on the external colon wall, affecting internal pressure, re-absorption, and healing.

Finally, fluid-structure interaction is foreseen by coupling active strain electromechanics with advanced fluid dynamics models, further modulating the intraluminal pressure profile and critically affecting the shear stress on the patch profile.

\section*{Acknowledgement}
Authors acknowledge the support of the Italian National Group for Mathematical Physics (GNFM-INdAM).

 \clearpage

\newpage
 \bibliographystyle{elsarticle-num-names} 
 \bibliography{cas-refs}

\clearpage

\newpage
\appendix
\section{Entraiment analysis of the ICC cells and mesh convergence analysis}
\label{sec:A}
Entrainment is one of the most significant mechanisms in gastrointestinal electrophysiology. This phenomenon is inherent. The excitability parameter and, consequently, the intrinsic frequency of electrical oscillations at the cellular level are larger in the upper than in the lower portion of the gastrointestinal tract. Excitability and frequency in the stomach are obviously higher in the proximal region than in the distal region near the pylorus. Similarly, the pylorus has a greater frequency than the duodenum, which has a higher frequency than the jejunum and we have the same behavior in the colon. This implies that the oscillation frequencies of the ICCs in these areas would differ greatly if they were isolated. However, the ICCs in the lower region are trained due to the cellular interaction at the tissue and organ level and for good movement coordination. To put it another way, the high-frequency cells and the low-frequency cells are compelled to vibrate at about the same frequency.\\

\begin{figure}[h] 
\centering
\includegraphics[scale = 0.7]{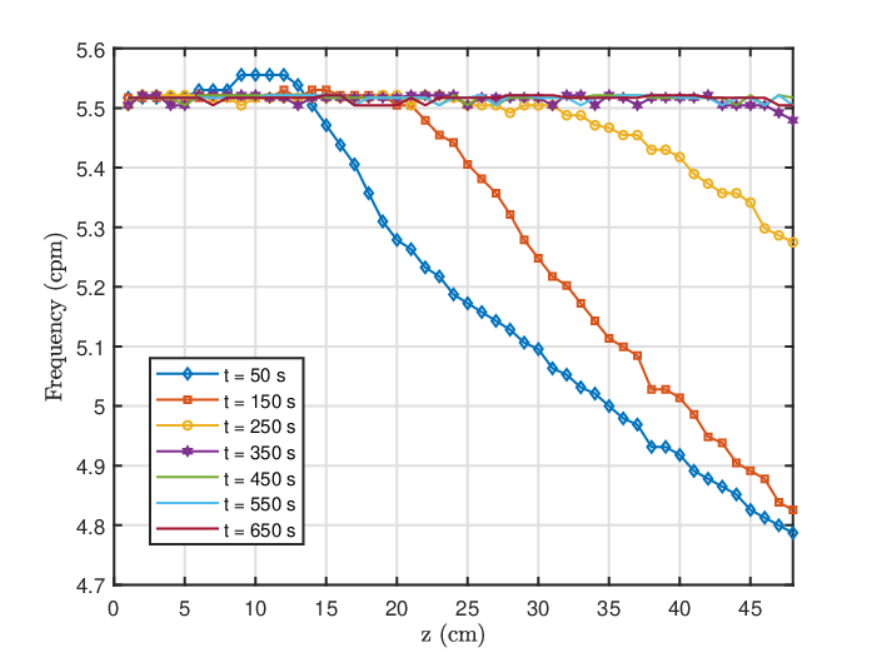}
\caption{Temporal evolution of the frequency of the ICC at each point}
\label{fig:A1}
\end{figure}

In accordance with this theory, we ran a simulation for 1000 s on a computational domaine. In order to determine the frequency, we then extracted the data along a line that extended from [0,50] cm. Fig .\ref{fig:A1} depicts the frequency gradient's temporal progression. Initially, every point exhibits its own frequency of oscillation. As a result, at $t = 0 s$, there is a noticeable frequency gradient throughout the domain. Gradient reduces with time. The electrical oscillations of the ICC demonstrate nearly constant frequency from $t = 550 s$, which is the same as the ICC's initial frequency in the top region of the domain (i.e. at $z = 0$).\\

\begin{figure}[h] 
\centering
\includegraphics[scale = 0.7]{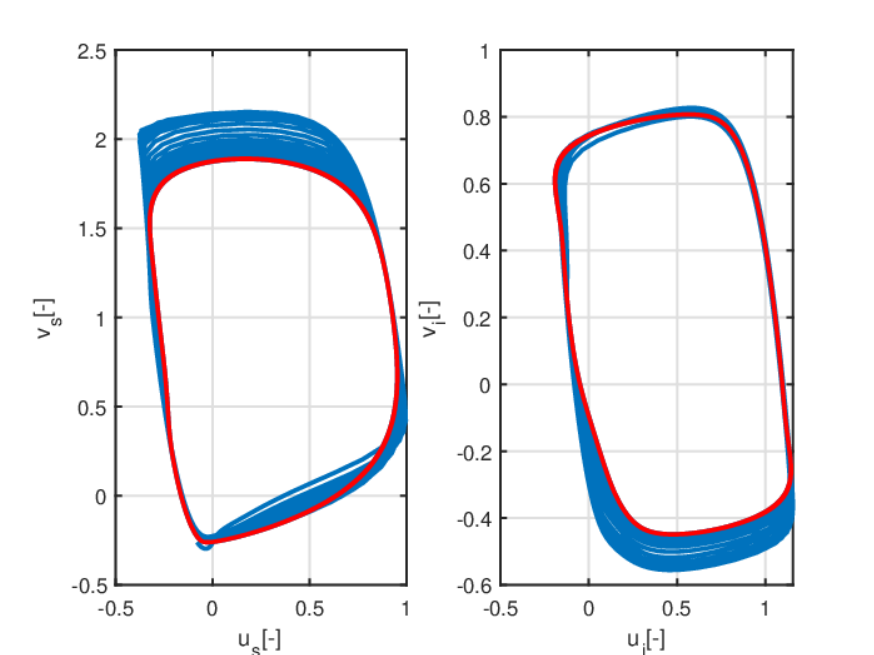}
\caption{Development of the phase portraits from the intrinsic to the entrained state of ICCs (right) and SMCs (left) located in the middle of the domain at $z = 25 cm$. The resulting stable limit cycle is shown in red.}
\label{fig:A2}
\end{figure}

This suggests that these pacemaker ICCs in the upper portion of the domain have driven the ICCs at the other end. Horizontal straight lines show that training from $t=550s$ is consistent throughout the domain. This outcome ensures that the model can replicate a physiological entrainment procedure. In order to validate this behavior, we present in Fig.\ref{fig:A2} the progression of the phase portrait for the two electrophysiological state variables for SMC and ICC from their intrinsic to entrained states. The last $450$ seconds, which we have highlighted in red, demonstrate how both cell types form a stable limit cycle. This indicates that, after 550 seconds, the system has stabilized. \citet{brandstaeter2018computational} was the first to perform this kind of analysis. This analysis's ability to solve the electrophysiological problem alone for up to $550$ seconds before linking it to the mechanics problem is a crucial feature. 
To assess the accuracy of the numerical scheme described in the previous section, a series of numerical tests is carried out. The first test aims to study the convergence of the electrophysiological solver towards the physical solution as a function of mesh size or degree of freedom (DoF).  For this purpose, we consider the same cylindrical domain as in the previous. For simplicity and to facilitate the implementation of this test, the domain is constructed directly in \texttt{FEniCs} and triangular Lagrangian finite elements and a homogeneous mesh of size  are used. Different refinements are considered based on the number of subdivisions  $ N \in [50, 100, 150, 200, 250, 300, 350, 400, 450]$. Each simulation runs for a total duration $t = 800 s$ with a constant time step $\Delta t = 0.1 \rm s$.\\

It is well known that the conduction velocity of reaction-diffusion systems generally depends on the numerical scheme. This analysis aims to obtain a physiologically acceptable conduction velocity (CV). \cite{quarteroni2017integrated}. \\
\begin{figure}[h] 
\centering
\includegraphics[scale = 0.7]{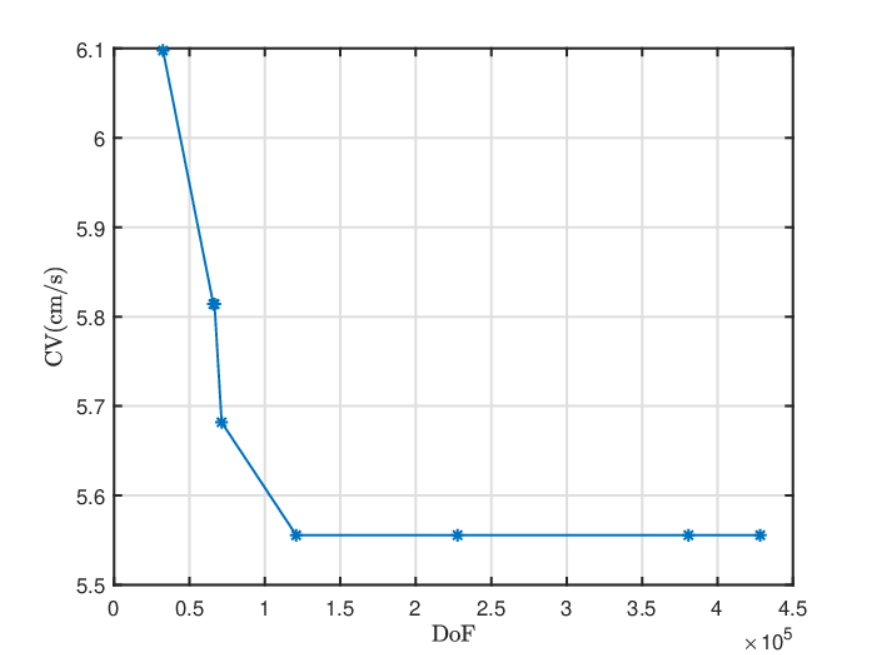}
\caption{conduction velocity analysis}
\label{fig:A3}
\end{figure}

We calculated the conduction velocity at a point located in the middle of the domain to avoid contamination of the boundary conditions. Fig.\ref{fig:A3} shows us the evolution of the conduction velocity as a function of the degree of freedom. we see that a coarse mesh tends to overestimate the conduction velocity. the more the number of degrees of freedom increases, the more the conduction velocity tends towards a stable value. This simulation also shows us that the conduction velocity is within the range of physiologically acceptable values \citep{sanders2014interstitial}.
\section{Tuning of mechanical parameters}
\label{sec:B}
To calibrate our passive material model and associated code, we undertook a series of tests to calibrate the model parameters. These tests include a uniaxial test based on Nagaraja data \citep{nagaraja2021phase}, and a triaxial test using cylinder occlusion based on Sokolis data \citep{sokolis2013microstructure}. We would like to point out that as we did not have the original experimental data, we were content to validate part of the authors' experimental curve. To this end, we have extracted data from their figure to use it as a reference. About the uniaxial test, most of the parameters were taken from \citet{nagaraja2021phase} study, which we then used as a basis for approximating certain points in \citet{sokolis2013microstructure} experimental data. \\

\begin{figure}[h!]
    \centering
    \includegraphics[scale=0.75]{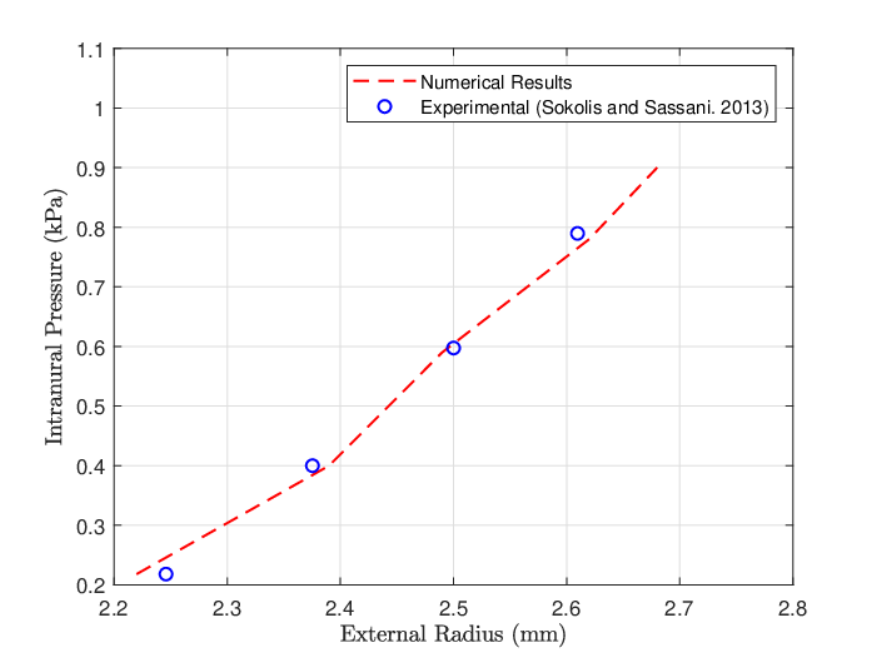}
    \caption{Comparison between simulation of the three-axial test and experimental data.}
    \label{fig:B2}
\end{figure}

We began by attempting to reproduce Nagaraja's experimental results with our model. To this end, we performed a uniaxial test aimed at reproducing Nagaraja's experimental results for the case where $\beta$ (the cutting angle of the sample.) is equal to $90^{\circ}$. This test was performed for the maximum and minimum values of the parameters, and some parameters were adjusted to approximate the experimental values. Figure \ref{fig:B1} below illustrates the comparison between the model and the experimental. We can see that the model comes closer to the experimental curve for the minimum parameters. This shows that these parameters can be used without causing errors in the code. The calibrated data are summarised in the following table\ref{table:T1}.\\
\begin{table}[h!]
\centering
\caption{Table of Material used for the uni-axial test \citep{nagaraja2021phase}}
\begin{tabular}{c c c c c c c c c} 
 \hline
sets & $\mu [kPa]$ & $k_1^l[kPa]$ & $k_2^l[-]$ & $k_1^c[kPa]$ & $k_2^c[-]$ & $k_1^d[kPa]$ & $k_2^d[-]$ & $\theta [\circ]$\\
 \hline
 max & $5$ & $77.35$ & $1.04$ & $0.95$ & $0.06$ & $7.38$ & $0.6$ & $39.78$\\
 min & $5$ & $5.14$ & $1.19$ & $0.78$ & $0.02$ & $3.65$ & $0.31$ & $38.18$\\
 \hline
\end{tabular}
\label{table:T1}
\end{table}
\begin{figure}[h]
    \centering
    \includegraphics[scale=0.75]{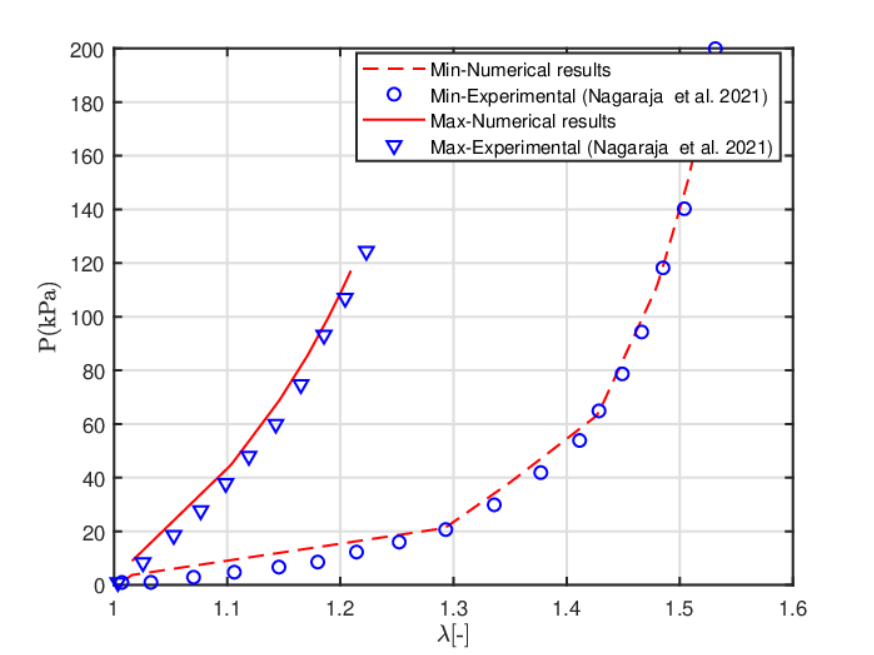}
    \caption{Comparison between simulation of the uni-axial test and experimental data.}
    \label{fig:B1}
\end{figure}
To be more consistent about the model's performance, we carried out a carefully calibrated inflation test. As shown in the figure \ref{fig:B2}, a $3.5$ $cm$ long cylinder was used for this purpose \citep{sokolis2013microstructure}. Pressure was applied to the inner surface of the cylinder, while both ends were clamped. After each inflation, the parameters were carefully adjusted to reproduce the experimental data faithfully and to extract the outer radius. The agreement between the model and the experimental data from \citep{sokolis2013microstructure} is clearly demonstrated in the figure below. The parameters obtained after calibration are detailed in Table \ref{table:T2}. These data will be crucial for all future simulations.
\begin{table}[h!]
\centering
\caption{Table of Material used for the three-axial test}
\begin{tabular}{c c c c c c c c c} 
 \hline
parameters & $\mu [kPa]$ & $k_1^l[kPa]$ & $k_2^l[-]$ & $k_1^c[kPa]$ & $k_2^c[-]$ & $k_1^d[kPa]$ & $k_2^d[-]$ & $\theta [\circ]$\\
 \hline
 values & $2.5$ & $5.4324$ & $1.19$ & $0.78$ & $0.02$ & $3.65$ & $0.31$ & $39.5$\\
 \hline
\end{tabular}
\label{table:T2}
\end{table}

\newpage
\section{Fibers generation procedure}
\label{sec:C}
\begin{figure}[h]
    \centering
    \includegraphics[scale=0.6]{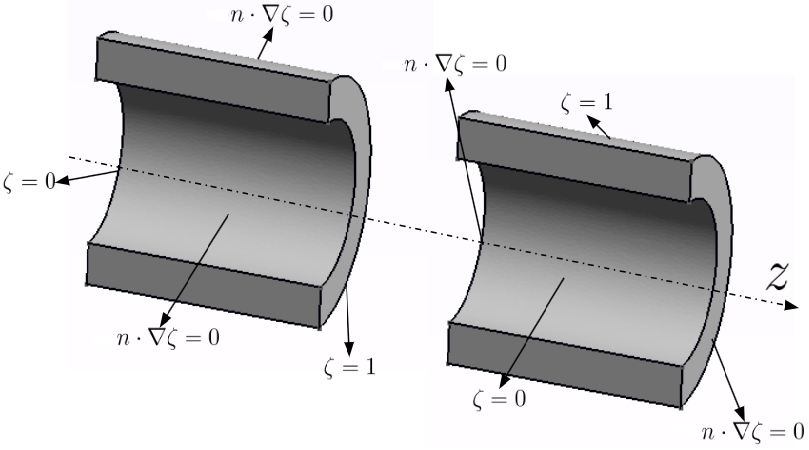}
    \caption{Configuration used to compute fiber orientation in the colon, left represents the configuration for the longitudinal fibers and right the configuration for the radial fibers.}
    \label{fig:C1}
\end{figure}

We start by the configuration show in Figs \ref{fig:C1}. The curvilinear coordinates are obtained by solving the stationary modified scalar diffusion  with the corresponding
boundary conditions.
\begin{subequations}
\begin{align}
\nabla^2 \zeta &=  0 \quad on \quad \Omega_0\\
\zeta &= \zeta_0 \quad on \quad \partial\Gamma_D \\
\bn\cdot \nabla \zeta &= 0 \quad on \quad \partial\Gamma_N
\end{align}
\label{eq:C1}
\end{subequations}
where $\zeta$ is an arbitrary scalar field, $\zeta_0$ is the value prescribed on the Dirichlet boundary ($\partial\Gamma_D$), and $\bn$ is the unit vector of the surface normal.
The discrete longitudinal vector field $\nabla \zeta_z$ is obtained by solving Eqs. \ref{eq:C1} with the Neumann boundary condition prescribed on the inner and outer surfaces, together with the Dirichlet boundary condition prescribed on both end surfaces as shown in Figs. \ref{fig:C1}. The final longitudinal fiber direction is obtained by the normalisation $\bn_l=\nabla\zeta_z/ \Vert  \nabla\zeta_z \Vert $. \\
The radial vector direction  $\bn_r$ is obtained similarly, only the Neumann and Dirichlet boundary conditions must be interchanged. Then, the circumferential fiber direction $\bn_c$ is defined as a cross product of the unitary radial and longitudinal vector fields, $\bn_r$ and $\bn_l$. Regarding the helical fiber direction, two additional parameters are necessary, a unit vector $\br_0$ aligned with the centreline and the angle $\theta$ that will determine the rotational anisotropy from the circumferential direction. Once the radial direction has been computed, we project the centreline on the normal direction and compute the so-called flat fiber field $\br_f$\citep{ruiz2020thermo}. Then the diagonal fibers are obtained by using the Rodrigues rotation formula.   
\begin{equation}
    \bn_{d1}= \br_f \cos(\theta)+ (\bn_l \cross \br_f)\sin(\theta) + \bn_l(\bn_l \cdot \br_f)(1-\cos(\theta))
\end{equation}
a similar expression can be derived for $\bn_{d2}$.
\newpage
\section{Zoomed clip of the region of
interest where the patch is located}
\label{sec:D}
\begin{figure}[h] 
\centering
\includegraphics[width=\textwidth]{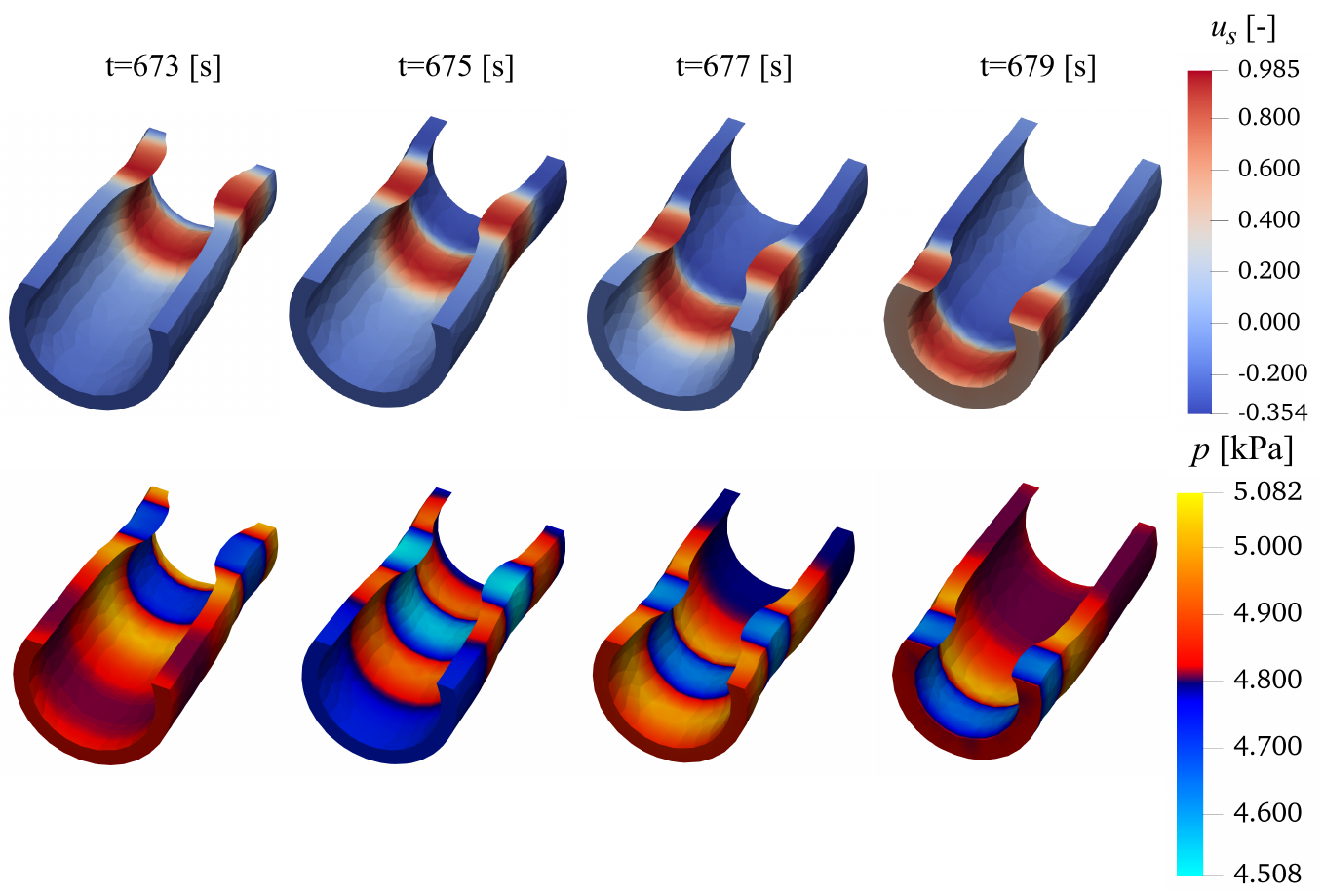}
\caption{Temporal evolution of the hydrostatic pressure $p$, the action potential in the smooth muscle layer $u_s$, and the longitudinal and circumferential fibers distribution in the deformed domain corresponding to healthy condition ($\mu_p=\mu_t$).}
\label{fig:4}
\end{figure}

\newpage
\section{Numerical solution of the nonlinear diffusivity with $D_s^p=0.01D_s$ and $D_i^p=0.01D_i$}
\label{sec:E}
\begin{figure}[ht!] 
\centering
\includegraphics[width=0.95\textwidth]{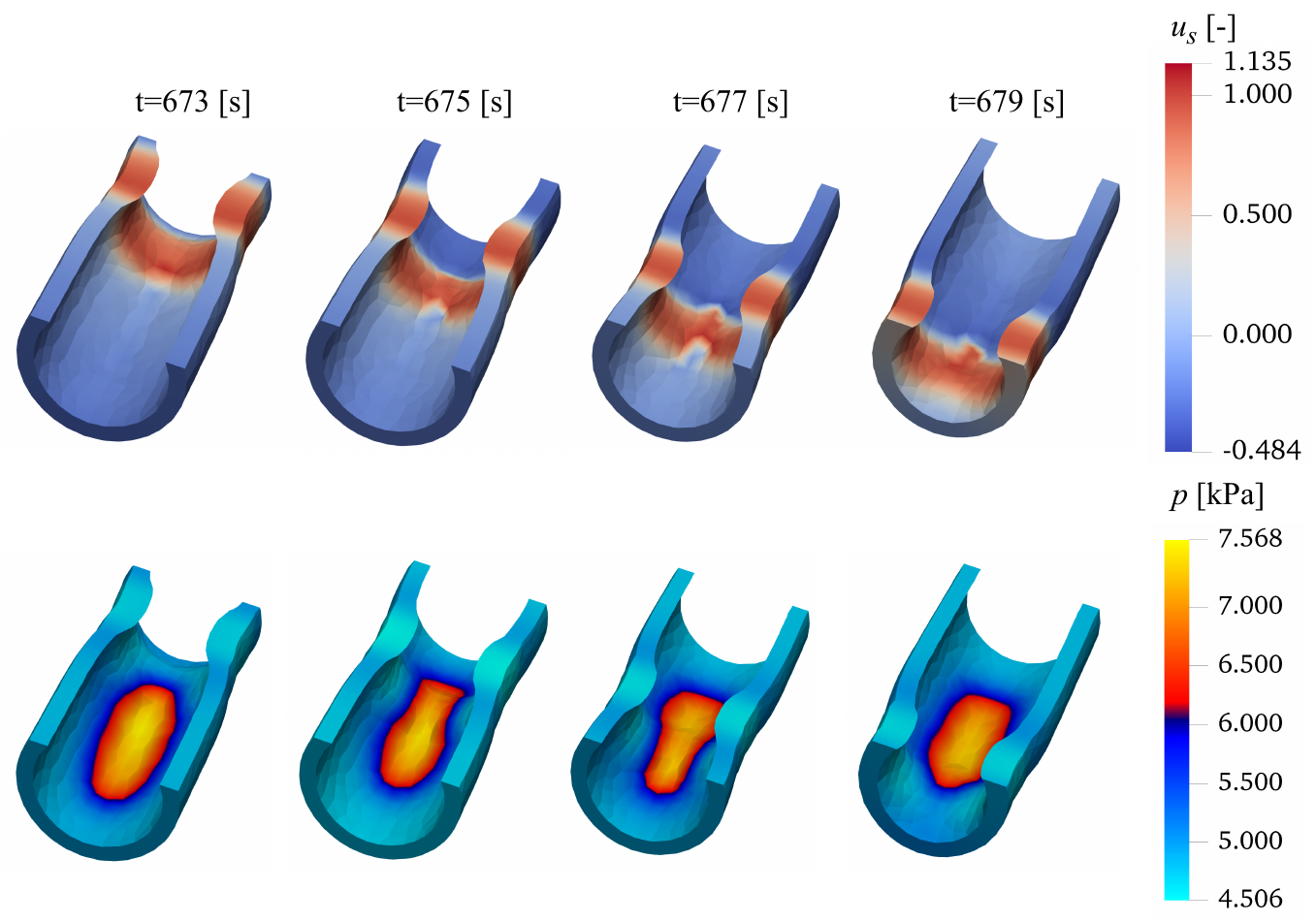}
\\
\includegraphics[width=\textwidth]{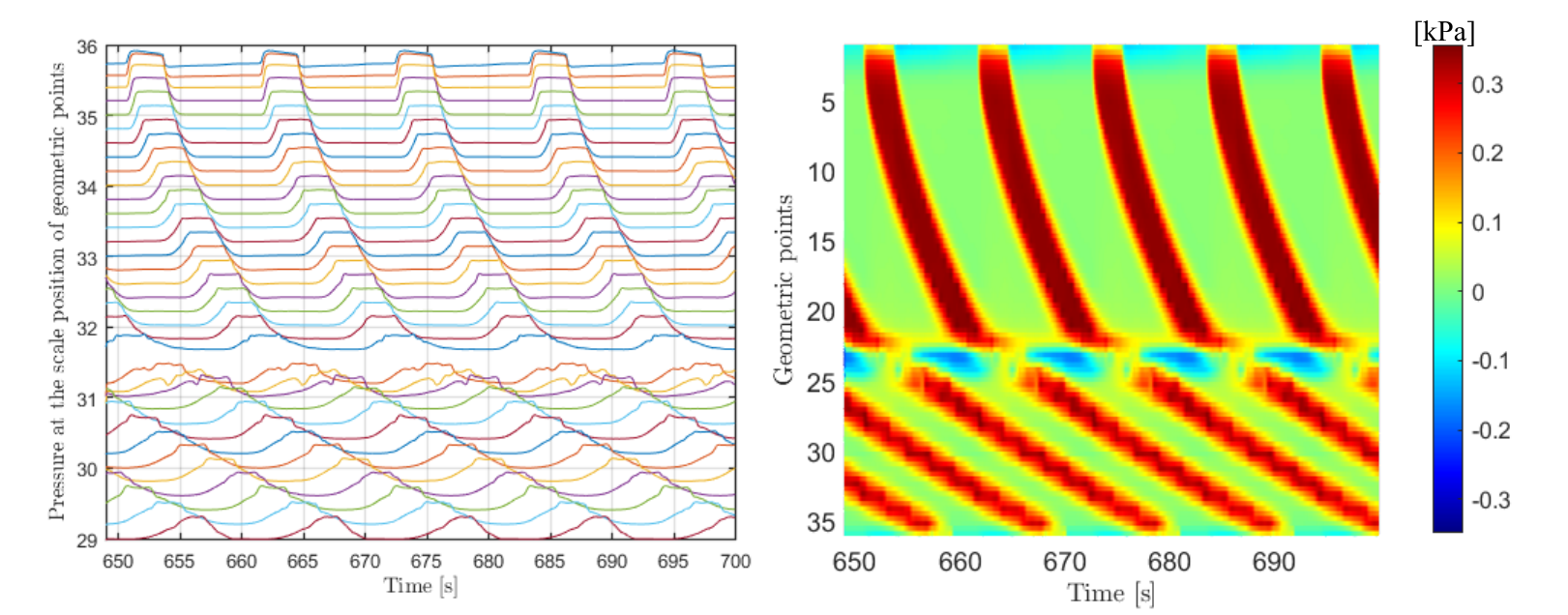}
\caption{Temporal evolution of hydrostatic pressure $p$ and the action potential in the smooth muscle layer $u_s$ in a region around an elliptical geometry of radii $r_{\rm max}= 3$ and $r_{\rm min}=2$ with stiffness $\mu_p=2 \mu_t$ and the diffusion coefficients $D_s^p=0.01D_s$ and $D_i^p=0.01D_i$.
HRM map  with $\mu_p=2 \mu_t$ with in-homogeneous diffusivity $D_s^p=0.01D_s$ and $D_i^p=0.01D_i$}
\label{fig:11E}
\end{figure}





\end{document}